\documentclass[12pt,preprint]{aastex}

\newcommand{\mach}{{\cal M}}
\newcommand{\lbox}{\mathrm{L_{box}}}

\newcommand{\Lam}{\Lambda}
\newcommand{\del}{\delta}

\newcommand{\hinv}{{\rm h^{-1}}}
\newcommand{\himpc}{\hinv{\rm\,Mpc}}
\newcommand{\hikpc}{\hinv{\rm\,kpc}}
\newcommand{\kms}{{\rm\,km\ s^{-1}}}
\newcommand{\kmsmpc}{{\rm\ km\ s^{-1}\ Mpc^{-1}}}
\newcommand{\Gyrs}{{\rm Gyrs}}
\newcommand{\yrs}{{\rm yrs}}
\def\H{{\rm H}}
\newcommand{\Msun}{M_{\odot}}
\newcommand{\vv}[1]{{\bf #1}}

\newcommand{\etal}{et~al.}
\newcommand{\ie}{{\frenchspacing i.e.}}
\newcommand{\eg}{{\frenchspacing e.g.}}
\newcommand{\ltsim}{\lesssim}
\newcommand{\gtsim}{\gtrsim}
\def\expec#1{\langle#1\rangle}

\def\bgeqa#1{\begin{eqnarray}\label{#1}}
\def\endeqa{\end{eqnarray}}

\def\Fig#1{Figure~\ref{#1}}

\newcounter{thetabs}
\newcommand{\tabnum}{\arabic{thetabs}}

\shorttitle{Cosmic Mach Number}
\shortauthors{Nagamine et al.}

\begin{document}

\title{Cosmic Mach Number as a Function of Overdensity \\
and Galaxy Age}


\author{Kentaro Nagamine}
\affil{Joseph Henry Laboratories, Physics Department, 
Princeton University, \\ 
Princeton, NJ 08544}

\author{Jeremiah P. Ostriker, and Renyue Cen}
\affil{Princeton University Observatory, Princeton, NJ 08544}
\email{(nagamine,jpo,cen)@astro.princeton.edu}

\begin{abstract}
We carry out an extensive study of the cosmic Mach number ($\mach$) 
on scales of $R=5, 10$ and $20\himpc$ using a LCDM hydrodynamical 
simulation. We particularly put emphasis on the environmental 
dependence of $\mach$ on overdensity, galaxy mass, and galaxy age.
We start by discussing the difference in the resulting $\mach$ 
according to different definitions of $\mach$ and different 
methods of calculation.
The simulated Mach numbers are slightly lower than the 
linear theory predictions even when a non-linear power spectrum
was used in the calculation, reflecting the non-linear evolution
in the simulation.
We find that the observed $\mach$ is higher than the simulated
mean $\expec{\mach}$ by more than 2-standard deviations, 
which suggests either that the Local Group
is in a relatively low-density region
or that the true value of $\Omega_m$ is $\sim 0.2$, 
significantly lower than the simulated value of 0.37. 
We show from our simulation that the Mach number is a 
weakly decreasing function of overdensity. 
We also investigate the correlations between galaxy age, 
overdensity and $\mach$ for two different samples of galaxies 
--- DWARFs and GIANTs. 
Older systems cluster in higher density regions with lower 
$\mach$, while younger ones tend to reside in lower density regions with 
larger $\mach$, as expected from the hierarchical structure 
formation scenario.
However, for DWARFs, the correlation is weakened by 
the fact that some of the oldest DWARFs are left over
in low-density regions during the structure formation history.
For giant systems, one expects blue-selected 
samples to have higher $\mach$ than red-selected ones.
We briefly comment on the effect of the warm dark matter 
on the expected Mach number.
\end{abstract}


\section{Introduction}
\label{introduction_section}
The cosmic Mach number ``$\mach$'' is the ratio of the bulk flow ``$V$'' of the  
velocity field on some scale $R$ to the velocity dispersion ``$\sigma$'' 
within the region. It was introduced by \citet[hereafter OS90]
{Ostriker90}, who stressed that it is independent of the normalization 
of the power spectrum,
and is insensitive to the bias between galaxies and dark matter (DM). 
Basically, it characterizes the warmth or coldness of the velocity 
field by measuring the relative strength of the velocities at scales 
larger and smaller than the patch size $R$, so that it effectively 
measures the {\em slope} of the power spectrum at the scale corresponding
to the patch size. OS90 made rough estimates of $\mach$ 
using available observational data on three different scales, 
and found that the observed $\mach$ was higher than the expected 
values of the standard cold dark matter model ($\Omega_m=1$ where $\Omega_m$ 
is the cosmological matter-density divided by the critical density of
the universe; hereafter SCDM) 
in the linear regime by more than a factor of 2 
(${\cal M}_{obs}\simeq 1-4$ and ${\cal M}_{SCDM}\ltsim 1$).
Subsequently, \citet{Suto90}, using N-body simulations, argued that 
the constraint on $\mach$ derived by OS90 holds at the 90$\%$ 
confidence level, and that the distribution of $\mach$ is close to 
Maxwellian in linear and mildly non-linear regimes. 
\citet{Park90} has also argued that the biased open CDM models
are preferred to the SCDM models using an N-body simulation.

The first serious calculation of $\mach$ using the first
generation of large-scale hydrodynamical simulations which include 
star formation was carried out by \citet[hereafter SCO92]{Suto92}. 
Using this type of simulation enables one to examine the velocity field
of galaxies and DM independently without an ad hoc 
assumption of bias between galaxies and DM. 
They used the patch size of $R=18$ and $40\himpc$,
and argued that there was no significant difference in $\mach$
between galaxies and DM, although the galaxies had somewhat
larger $\sigma, V$ and $\mach$ than did DM. 
Their best estimate of the mean Mach number 
derived from SCDM simulations is $\langle \mach \rangle=0.6$, lower than 
the observational estimate of $\mach_{obs} \gtsim 1$. 

\citet[hereafter S93]{Strauss93} made more realistic and 
direct comparison of observations and models.
Accepting the fact that the existing peculiar velocity data do not 
allow us to compute the ideally defined $\mach$ as in OS90, they 
defined a modified Mach number which incorporates the observational
errors in measured distances due to the scatter in the Tully-Fisher relation.
They constructed a mock catalog of the observations using 
SCDM hydrodynamical simulations similar to those that were used by SCO92,
and calculated their modified $\mach$ from them.
S93 obtained smaller $\mach$ than did OS90 because they included all
velocity components on scales less than the bulk flow into 
the velocity dispersion,
whereas OS90 erased the small-scale dispersion by smoothing.
As a consequence, the estimates of S93 on $\sigma$ is larger, 
resulting in a smaller $\mach$.
They found that 95\% of the mock catalogs had smaller $\mach$ 
than observed, and that the Mach number test rejects the 
SCDM scenario at 94\% confidence level.

Since $\mach$ is defined as $V/\sigma$ on a certain scale $R$, 
a larger $\mach$ implies a smaller $\sigma$ if the variation of $V$ 
is weaker than that of $\sigma$. 
Observationally, it has been recognized for at least a decade that 
the velocity field is very cold outside of clusters
\citep{Brown87, Sandage86, Groth89, Burstein90, Willick97, Willick98}.
We ask ourselves in this paper how typical such a cold region of space
would be in the entire distribution of the velocity field.
We note that \citet{Weygaert99} also address the same question 
using N-body simulations which simulate the Local Group 
via constrained initial conditions.

A closely related quantity is 
the pairwise velocity dispersion $\sigma_{12}$, which has been 
much studied due to its cosmological importance in relation to the 
``Cosmic Virial Theorem''. This theorem relates 
$\sigma_{12}$ to the two- and three-point correlation functions and $\Omega_m$.
Unfortunately, the determination of $\sigma_{12}$ is quite
unstable and its value differs significantly
from author to author \citep{Mo93, Zurek94, Somerville97, Guzzo97, Strauss98}. 
This is because $\sigma_{12}$ is a pair-weighted statistic and is  
heavily weighted by the objects in the densest regions. Inclusion or exclusion
of even $\sim 10$ galaxies from the Virgo Cluster can change
$\sigma_{12}$ by $\sim 100 - 200\kms$, and the correction for
the cluster-infall also affects the result significantly.

To overcome this problem, alternative statistics have been suggested. 
\citet{Kepner97} proposed the redshift dispersion as a new statistic, and 
suggested a calculation of the dispersion as a function of local 
overdensity $\delta$. They analytically showed that $\sigma_{12}$ 
is heavily weighted by the densest regions of the sample. 
In the same spirit, \citet{Strauss98} defined and calculated 
a new measure of $\sigma_{12}$ as a function 
of $\delta$ in redshift space using the Optical Redshift Survey
\citep{Santiago95}, and showed that $\sigma_{12}$ is indeed an increasing function 
of $\delta$.
As we will show in this paper, the above statement for $\sigma_{12}$ 
holds for the velocity dispersion $\sigma$ as well; it is 
heavily weighted by the densest regions.
\citet{Davis97} proposed a single-particle-weighted statistic
which measures the one-dimensional rms peculiar velocity dispersion 
of galaxies, and applied the statistic to the sub-sample of
the Optical Redshift Survey and the 1.2 Jy $IRAS$ catalog \citep{Fisher95}.
\citet{Baker00} applied the same statistic to the Las Campanas
Redshift Survey \citep{Shectman96}, and find that the low-$\Omega_m$
is favored when the results are compared with N-body simulations. 
We also note the work by \citet{Juszkiewicz99} who derived a simple 
closed-form expression relating the mean pairwise relative velocity $v_{12}$
to the two-point correlation function of mass. Their results 
can also be used to estimate $\Omega_m$. 

Since the original work of OS90, many things have changed.
The resolution and the accuracy of simulations have increased 
significantly due to the increased computer power
and more realistic modelling of galaxy formation. 
The favored cosmology shifted from SCDM to LCDM ($\Lam$-dominated 
flat cold dark matter model), as more and more modern 
observational data suggest a flat low-$\Omega_m$ universe with a  
cosmological constant $\Lam$ \citep[\eg,][]{Efstathiou90, Ostriker95, 
Turner97, Perlmutter98, Garnavich98, Bahcall99, Balbi00, Lange00, Hu00}.
Thus, we are motivated to study this statistic again using 
a state-of-the-art LCDM hydrodynamical simulation, 
which allows us to treat baryons and dark matter separately
without invoking an ad hoc bias parameter.

In this paper, we calculate $\sigma, V$, and $\mach$
with patches of size $R=5, 10$ and $20\himpc$ using an 
LCDM hydrodynamical simulation which includes 
star formation. The details of the simulation are explained in 
\S~\ref{simulation_section} and in Appendix A.
We correct for the underestimation of the bulk flow due to the 
limited size of the simulation box using the linear theory
of gravitational instability in \S~\ref{theory_section}.
In \S~\ref{method_section}, we describe the method of the calculation of 
$V$, $\sigma$, and $\mach$.
The results of the calculation is presented in \S~\ref{result_section},
where we discuss the difference in $\mach$ resulting from 
different methods of calculation. 
The distribution of $\mach$ is discussed in \S~\ref{distribution_section}.
In \S~\ref{overdensity_section} and \ref{age_section}, we divide the sample of 
simulated galaxies by the local overdensity and by their age, 
and study the correlation with $\mach$
under the hierarchical structure formation scenario.

All velocities in this paper are presented in
the CMB-frame, and the velocity dispersion is 3-dimensional
(\ie, not just the 1-dimensional line-of-sight component).

\section{The Simulation}
\label{simulation_section}
The hydrodynamical simulation we use here is similar to but greatly 
improved over that of \citet{CO92a, CO92b}. 
The adopted cosmological parameters are $\Omega_m=0.37$, $\Omega_{\Lambda}=0.63$, 
$\Omega_b=0.049$, $n=0.95$, $\sigma_8=0.8$ and $h=0.7$, 
where $H_0=100h\kmsmpc$and $n$ is the primordial power spectrum index. 
The power spectrum includes a 25$\%$ tensor mode contribution to 
the cosmic microwave background fluctuations on large scales.
The present age of the universe with these parameters is 12.7 \Gyrs.
The simulation box has a size of $\lbox$=100$\himpc$ and $512^3$ grid points, 
so the comoving cell size is $200\hikpc$.
It contains $256^3$ dark matter particles, each weighing 
$5.3\times10^9\hinv\Msun$. 

The code is implemented with a star formation recipe summarized in 
Appendix A.
It turns a fraction of the baryonic gas in a cell into a collisionless particle 
(hereafter ``galaxy particle'') in a given timestep 
once the following criteria are met simultaneously (see Appendix A): 
1) the cell is overdense, 
2) the gas is cooling fast, 
3) the gas is Jeans unstable, 
4) the gas flow is converging into the cell. 

Each galaxy particle has a number of attributes at birth, including 
position, velocity, formation time, mass, and initial gas metallicity.
Upon its formation, the mass of the galaxy particle is determined by 
$m_{\ast}=c_{\ast} m_{gas}\Delta t/t_{\ast}$, where $c_{\ast}$ is 
the star formation efficiency parameter which we take to be $c_{\ast}=0.25$.
$\Delta t$ is the current time-step in the simulation and $t_\ast=\mathrm{max}
(t_{\mathrm{dyn}}, 10^7\yrs)$.
The galaxy particle is placed at the center of the cell after its 
formation with a velocity equal to the mean velocity of the gas,
and followed by the particle-mesh code thereafter as collisionless 
particles in gravitational coupling with DM and gas. 
Galaxy particles are baryonic galactic subunits with masses ranging 
from $10^3$ to $10^{10}\Msun$, therefore, 
a collection of these particles is regarded as a galaxy. 
Feedback processes such as ionizing UV, supernova energy, 
and metal ejection are also included self-consistently. 
Further details of these treatments can be found in \citet{CO92a, CO92b}.
We also refer the interested readers to 
\citet{CO99a, CO99b, CO00, Blanton99, Blanton00, Nagamine99, Nagamine00} 
where various analyses have been performed using the same simulation.

 In addition to the above $\lbox$=100$\himpc$ simulation, we have a newly 
completed $\lbox$=25$\himpc$ simulation with 6 times better spatial 
resolution and 260 times better DM mass resolution. 
We use this new simulation as a reference by verifying
that the same trend found in $\lbox$=100$\himpc$ simulation 
is seen in $\lbox$=25$\himpc$ simulation as well, 
although the velocity field in the new one 
is significantly underestimated as the box size is not large enough.

\section{Linear Theory and Definitions of $\mach$}
\label{theory_section}
Under the linear theory of gravitational instability, the mean square 
bulk flow and the mean square velocity dispersion in a window of size $R$ 
can be calculated as follows \citep[\eg,][OS90]{Peebles93}:

\bgeqa{eq1}
\expec{V^2(R)} &=& \frac{\Omega^{1.2}\H^2}{2\pi^2}
\int_{0}^{\infty}P(k)W^2(kR) dk \\
\expec{\sigma^2(R)} &=& \frac{\Omega^{1.2}\H^2}{2\pi^2}
\int_{0}^{\infty}P(k)[1-W^2(kR)] dk
\endeqa
where $P(k)$ is the power spectrum of density fluctuations, 
and $W(kR)$ is the Fourier
transform of the window function of size $R$. In this paper,
we adopt the tophat window function 
$W(x)=3(\sin x-x\cos x)/x^3$. 
The effect of the cosmological constant on the term 
$\Omega^{1.2}$ is small \citep{Lahav91, Martel91}. Although OS90 include an
observational correction term $R\,\nabla\cdot\vv{v}/3$ in the integrand,
we do not include this term since it makes only slight difference. 
The root mean square (rms) cosmic Mach number can be defined in two ways,
depending on how one does the averaging (OS90):
\bgeqa{eq3}
\expec{\mach^2(R)}^{1/2} = \left\langle \frac{V^2(R)}{\sigma^2(R)} \right\rangle^{1/2}
\hspace{0.5cm} {\mathrm or} \hspace{0.5cm}
\left(\frac{\langle V^2(R) \rangle}{\langle \sigma^2(R) \rangle}\right)^{1/2}.
\endeqa

In practice, we can calculate $V(R)$ and $\sigma(R)$ 
for each patch we take in the simulation, 
and assign $\mach(R)=|V(R)/\sigma(R)|$ to each patch. We can then later
take the ensemble average by 
\bgeqa{eq4}
\expec{\mach(R)} = \left\langle \left| \frac{V(R)}{\sigma(R)}\right| \right\rangle.
\endeqa
This method allows us to observe the distribution
of the Mach number before we take the ensemble average. 
In the next section, we show the differences between the 
different definitions. 

\begin{figure}
\plotone{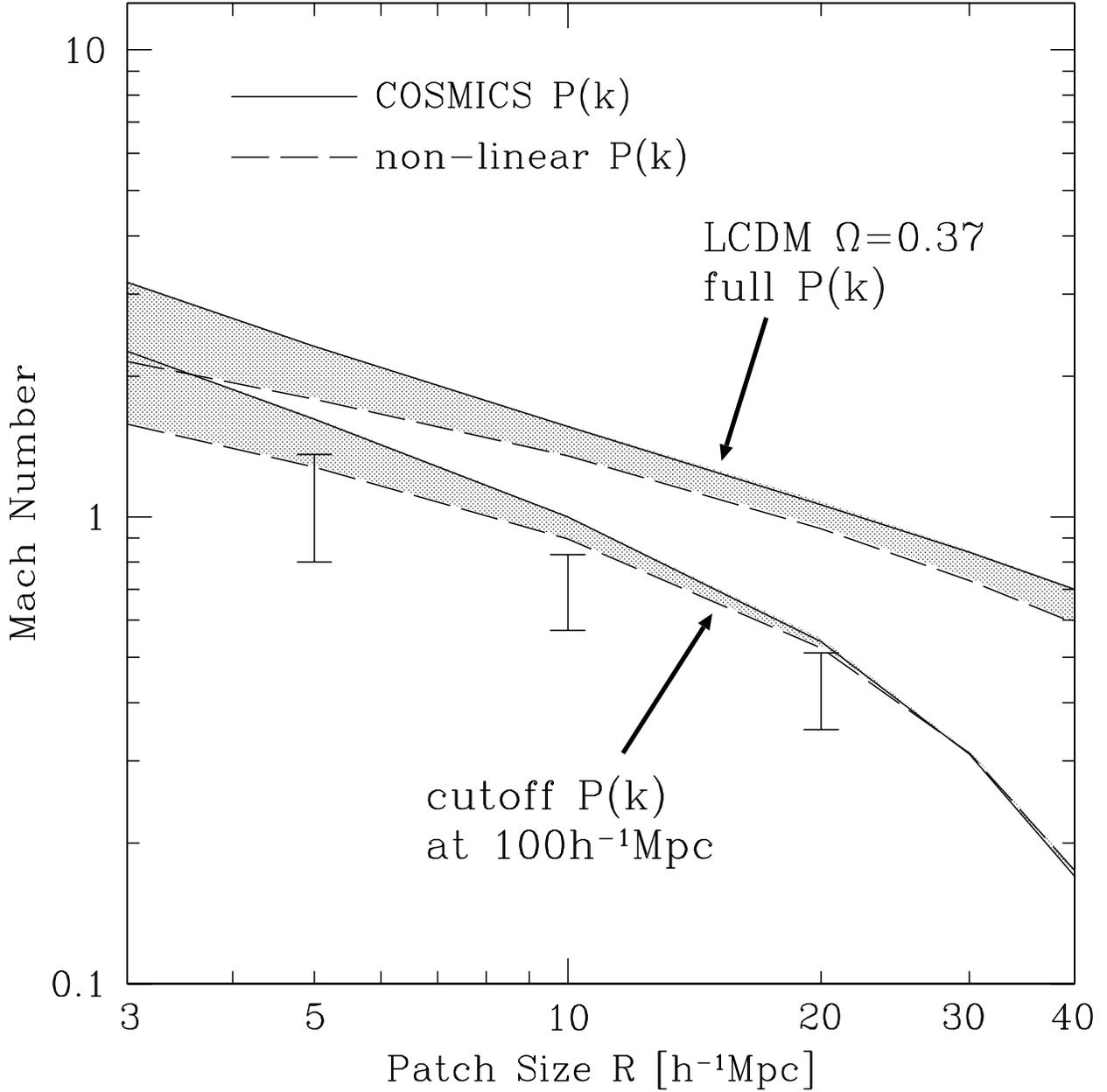}
\caption{
Cosmic Mach number as a function of scale $R$. The solid 
(top boundaries of the grey regions) and the dashed lines 
(bottom boundaries of the shaded regions) are the linear theory predictions 
calculated from the equations in \S~\ref{theory_section} using the
COSMICS $P(k)$ (solid line) and the non-linear $P(k)$ (short-dashed).
The non-linear $P(k)$ was evolved from an empirical 
double-power-law linear spectrum, and is known to provide a good fit to the observed 
optical galaxy power spectrum \citep{Peacock99}.
The top two lines are calculated using the full $P(k)$, and the 
bottom two from the cutoff $P(k)$ at $100\himpc$ to show the effect
of the limited simulation box size.
The three vertical lines indicate the range of simulated
raw values of $\mach$ before the correction of the bulk flow
for the lack of long wavelength perturbations. 
See \S~\ref{theory_section} and \S~\ref{method_section} for discussion.
\label{mach_scale_plain.ps}
}
\end{figure}

The simulation we use has a box size of only $\lbox$=100$\himpc$, 
and lacks long wavelength perturbations beyond this scale.
This lack of long wavelength perturbations results in an underestimate
of the bulk flow, as it is determined by the perturbations 
on scales larger than the patch size $R$. 
In particular, for the currently popular low-$\Omega_m$ models,
the peak of the power spectrum lies at scales larger than $200\himpc$.
So, a box size larger than $500\himpc$ is necessary for 
the correct and direct treatment of the bulk flow accurate to 10\% 
in the simulation on the scale of $20\himpc$. 
However, the rms bulk flow can be 
calculated correctly by the above equation at large enough scales.
The solid and the dashed lines in \Fig{mach_scale_plain.ps} 
show the predicted rms Mach number calculated from 
Equations~\ref{eq1}, 2, and \ref{eq3} (the latter definition).
The solid line is calculated by using the $P(k)$ obtained 
from the COSMICS package \citep{Bertschinger95} which was also used 
to generate the initial conditions of our simulation. 
The dashed line was calculated with the $P(k)$ which was
evolved to non-linear regime by \citet{Peacock96} scheme
from the empirical double-power-law linear spectrum. 
This non-linear $P(k)$ is known to provide a good fit to the
observed optical galaxy power spectrum \citep{Peacock99}. 
The two upper lines are calculated with the full $P(k)$, and 
the bottom two was calculated with the truncated $P(k)$ at 
$R=100\himpc$ to show the effect of the limited box size.
The non-linear $P(k)$ has more power on small-scales, therefore, 
predicts smaller $\mach$ than the linear $P(k)$ due to 
larger velocity dispersion. The scale-dependence of the predicted $\mach$ 
for the case of the full linear $P(k)$ can be well fitted by 
a power-law $\mach \propto R^{-0.6}$; {\it the cosmic Mach number 
is a decreasing function of scale $R$}.
The slope becomes shallower than this on smaller scales 
($R\ltsim 10\himpc$) for the non-linear $P(k)$ case.
We can also obtain the dependence on $\Omega_m$ of $\mach$ by 
calculating the linear theory prediction with different values
of $\Omega_m$ for the the full linear COSMICS $P(k)$. 
We obtain $\mach \propto \Omega_m^{-0.8}$ on the scale of 
$R=5-20\himpc$ for $0.2 \ltsim \Omega_m \ltsim 0.4$. 
The power index steepens to $-1.2$ for smaller values of $\Omega_m$, 
and gets shallower to $-0.7$ for $0.4 \ltsim \Omega_m \ltsim 1.0$.

\notetoeditor{Place \Fig{mach_scale_plain.ps} here.}

The three vertical lines in \Fig{mach_scale_plain.ps} 
at $R=5, 10$, and $20\himpc$ indicate
the range of rms values of the simulated $\mach$ 
for different methods of calculations summarized in Table~\ref{table1}.
On all scales, many of the simulated $\mach$ are smaller than 
the theoretical prediction by a factor of about 1.5, but the 
highest value in each case is consistent with the predicted value
with non-linear $P(k)$. The source of this slight discrepancy between 
the simulated and the predicted is probably due to the use of 
the linear theory equations, \ie, the non-linear effects are not
completely described by just plugging the non-linear $P(k)$ into 
the linear theory equations.

\begin{figure}
\plotone{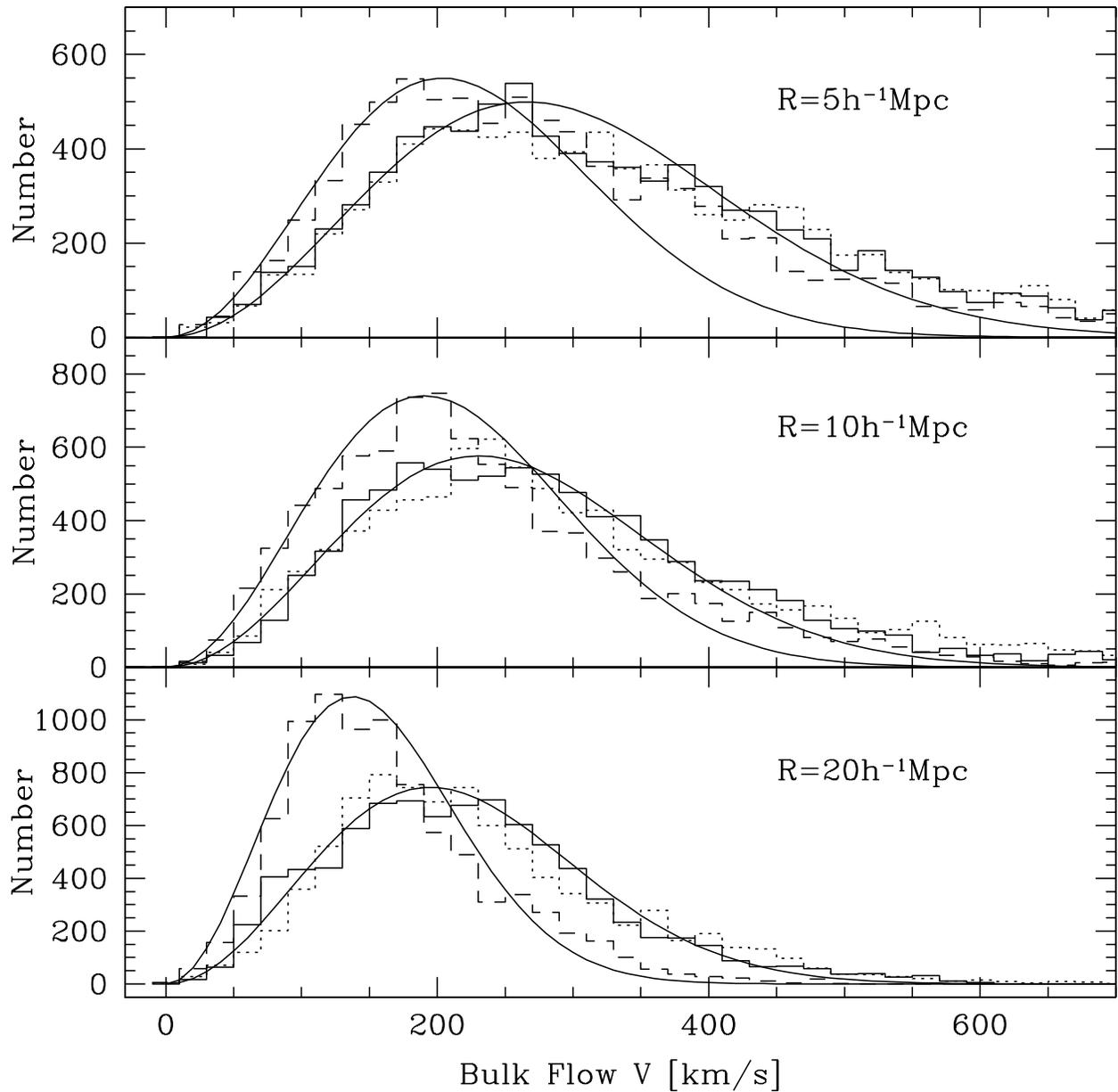}
\caption{
Bulk flow distribution of the simulated galaxies. 
The raw simulated bulk flow is shown by 
the dashed histogram. The solid histogram is the one after the
addition of the random Fourier components. The dotted histogram 
is obtained by simply multiplying numerical factors of 
1.2 ($R=5\himpc$), 1.25 ($R=10\himpc$), and 1.4 ($R=20\himpc$) to 
the raw simulated bulk flow.
The smooth curves are the `eye-ball' fits to the histograms by
Maxwellian distribution. See text for discussions.
\label{bulk_dist.ps}
}
\end{figure}

We wish to correct our simulated values of $V$ and $\mach$ for the
lack of long wavelength perturbations, but this is not a trivial task
\citep{Strauss95, Tormen96}. We first followed the method of \cite{Strauss95},
and computed the additional contribution to the bulk flow 
from the long wavelength perturbations larger than the simulation box size
by adding random phase Fourier components in Fourier space using
the linear theory equations.
In Figure~\ref{bulk_dist.ps}, we show the distributions of the simulated 
bulk flow before and after this process, calculated with the grouped 
galaxy velocities. The raw simulated bulk flow is shown by 
the short-dashed histogram. The solid histogram is the one after the
addition of the random Fourier components. The dotted histogram 
is obtained by simply multiplying the numerical factors of 
1.2 ($R=5\himpc$), 1.25 ($R=10\himpc$), and 1.4 ($R=20\himpc$) to 
the raw simulated bulk flow.
The smooth curves are the `eye-ball' fits to the histograms by
Maxwellian distribution.
All histograms show a good fit to the Maxwellian distribution
except that the raw simulated histogram of the $R=5\himpc$ has a longer 
tail than Maxwellian. 

\notetoeditor{Place \Fig{bulk_dist.ps} here.}

We find that the change in the distribution is fairly well approximated by 
simply multiplying a numerical factor to the raw simulated bulk flow.
We also confirm that the distribution does not change very much on 
the $\mach - V$ plane when
the random Fourier components of the bulk flow is added. 
Another thing is that the method of \citet{Strauss95} is explicitly
dependent on the normalization of the power spectrum.
On the other hand, if we simply take the ratio of the two rms Mach numbers
calculated with the full $P(k)$ and the cutoff $P(k)$ (the two 
solid or dashed lines in Figure~\ref{mach_scale_plain.ps}),
and use this ratio, we can correct the bulk flow being independent of 
the normalization of the power spectrum, though within 
the limitation of using the equations of the linear theory.

For these reasons, we choose to correct
for the lack of long wavelength perturbations in the latter manner,
as it is sufficient for our purpose.
The ratio of the two solid lines (COSMICS $P(k)$ case) in 
Figure~\ref{mach_scale_plain.ps} are 1.43 ($R=5\himpc$), 1.56 ($R=10\himpc$), 
and 1.96 ($R=20\himpc$). 
For the dashed lines (non-linear $P(k)$ case), the ratios are
slightly smaller; 1.30 ($R=5\himpc$), 1.50 ($R=10\himpc$), and 1.80 ($R=20\himpc$).
These factors are larger than the factors obtained by 
adding the random Fourier components. However, even if these correction
factors turn out to be overestimates, our conclusion will strengthen
in that case, because our corrected $\mach$ are still well below the observed $\mach$.
Hereafter, we adopt the correction factors of 1.43, 1.56 and 1.96 for  
$R=5, 10$ and 20$\himpc$ cases, respectively.

\section{Method of Calculation of $V$, $\sigma$, and $\mach$}
\label{method_section}
In this section, we describe how we calculate the bulk flow, 
the velocity dispersion, and the cosmic Mach number from our simulation.
We explore various options of calculations to see if they cause 
any difference in $\mach$. We are also interested in the difference
in $\mach$ of different tracers of the velocity field.

There are many ways one can place the patches in the simulation.
One also has to decide whether to use the particle-based
ungrouped data set, or to apply a grouping algorithm and identify 
galaxies and dark matter halos. 
Here, we consider the following cases:
\begin{enumerate}
\item{Particle-based:}
\begin{enumerate}
\item{centered on grouped galaxies: use galaxy particles (gal-pt)}
\item{centered on grouped galaxies: use DM particles (dm-galctr-pt)}
\item{centered on grouped DM halos: use DM particles (dm-dmctr-pt)}
\end{enumerate}
\item{Group-based:}
\begin{enumerate}
\item{centered on grouped galaxies: use grouped galaxy velocity (gal-gp)}
\item{centered on grouped DM halos: use grouped DM halo velocity (dm-gp)}
\end{enumerate}
\end{enumerate}

\begin{figure}
\plotone{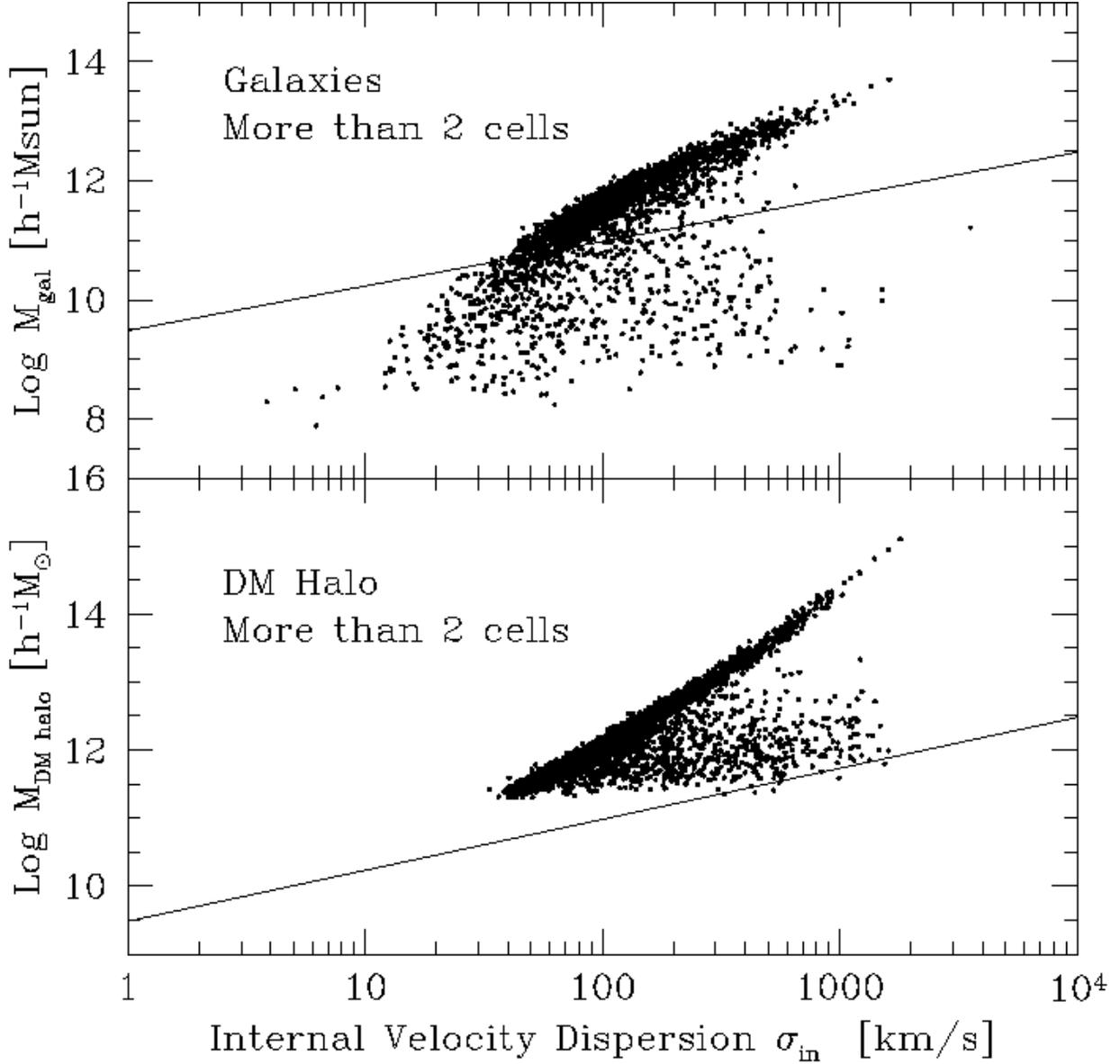}
\caption{
Mass of the grouped objects as a function of their internal velocity 
dispersion is shown for those objects which occupy more than 2 cells 
in the simulation. The solid line shows the cutoff boundary of 
$M_{group} \geq 3\times 10^9 \hinv\Msun \sigma_{in}^{3/4}$
for selecting out the dynamically stable objects as the center of the 
spherical tophat patches, 
where $M_{group}$ is the mass of the grouped object, and $\sigma_{in}$
is the internal velocity dispersion.
We pick those which lie above this boundary line as the centers of
the spherical tophat patches. DM halos are not affected by this cutoff.
\label{cutoff.ps}
}
\end{figure}

We first identify galaxies and DM halos 
in the simulation using the HOP grouping algorithm 
\citep{Eisenstein98}. Using a set of standard parameters
[$N_{dens},N_{hop},N_{merge}$]($\del_{peak},\del_{saddle},\del_{outer})
=[64,16,4](240,200,80)$, we obtain 8601 galaxies and 9554 DM halos in 
the simulation box. 
To select out dynamically stable objects as the centers of the patches,
we pick objects which occupy more than 2 cells in the simulation, 
and those which satisfy the criteria of 
$M_{group} \geq 3\times 10^9 \hinv\Msun \sigma_{in}^{3/4}$, where 
$M_{group}$ is the mass of the grouped object, and 
$\sigma_{in}$ is the internal velocity dispersion in units of $\kms$. 
This cutoff is motivated by looking at Figure~\ref{cutoff.ps} where
grouped objects which occupy more than 2 cells in the simulation 
are shown. DM halos are not affected by the latter cutoff.
We have confirmed that the results are robust to this pruning. 
We are left with 1585 galaxies and 4142 DM halos after this pruning. 
Changing the grouping parameters certainly affects the number of 
objects, which in turn affects the estimate of the velocity dispersion. 
Without grouping, for example, the velocity dispersion would be
over-estimated, as it would include the internal motions of particle
in each object. However, \citet{Eisenstein98} showed that the sample
is quite stable to the choice of parameters, so this effect is 
likely to be small. 
But this is an unavoidable numerical uncertainty and one should keep 
this in mind upon reading the results below.

\notetoeditor{Place \Fig{cutoff.ps} here.}

For the particle-based calculation, we calculate
the bulk flow $\vv{V}$ and the velocity dispersion $\sigma$ for 
each tophat patch in a mass-weighted manner:
$\vv{V}=\Sigma_j m_{p,j} \vv{v}_{p,j} / \Sigma_j m_{p,j}$
and 
$\sigma^2 = \Sigma_j m_{p,j}(\vv{v}_{p,j}-\vv{V})^2/ \Sigma_j m_{p,j}$,
where $\vv{v}_{p,j}$ and $m_{p,j}$ are the particle velocity and mass, 
and the sum $\Sigma_j$ is over all particles in the spherical tophat patch
of a given radius. 

For the group-based calculation, we need to calculate the mean velocity
of each grouped galaxy and DM halo first.
The mass-weighted mean velocity (center-of-mass velocity) of the $i$-th 
object $\vv{v}_i$ is calculated by
$\vv{v}_i = \Sigma_j m_{p,j} \vv{v}_{p,j} / \Sigma_j m_{p,j}$, 
where the sum is over all particles associated with the object. 
We call this velocity $\vv{v}_i$ as the galaxy velocity
or the DM halo velocity.
We then place spherical tophat patches of radius $R=5, 10$ and 
$20\himpc$ at the centers of the grouped objects, and calculate 
$V$ and $\sigma$ for each patch using objects' velocity $\vv{v}_i$ for 
galaxies and DM halos separately:
$\vv{V}=\frac{1}{N} \Sigma_i \vv{v}_i$ 
and 
$\sigma^2 = \frac{1}{N-1} \Sigma_i (\vv{v}_i - \vv{V})^2$,
where $N$ is the number of objects in the patch and the 
sum $\Sigma_i$ is over all the objects in the patch.  
Note that we do not weight by the mass in the group-based calculation
to mimic the real observations of galaxies.
Periodic boundary conditions are used for all the calculations.

All the calculations are done in real space as it is more
straightforward than doing it in Fourier space. 
We did not smooth the velocity field prior to these calculations.
The effect of the smoothing is discussed in OS90, where they 
noted that the non-zero smoothing length simply increases 
the theoretical prediction of $\mach$ compared with the non-smoothed
case. This is obvious because smoothing would erase the velocity 
dispersion on scales smaller than the smoothing length. 
Here, our intention is not to erase the small-scale 
dispersion by the smoothing, rather, to observe it as a function of 
local overdensity.

\section{Results}
\label{result_section}

\subsection{Mean and rms of $V$, $\sigma$, and $\mach$}
\label{mean_section}
From above calculations, we now have $V\equiv|\vv{V}|$ and $\sigma$ 
for each patch. We can now calculate the mean and rms Mach number  
following Equations \ref{eq3} and \ref{eq4} for both galaxies and DM. 
We summarize the results in Table~\ref{table1}.

\notetoeditor{Place Table~\ref{table1} here.}

The standard deviation (SD) is indicated to show the typical uncertainty
associated with the calculation of the mean in each case,
although the error in the mean is not exactly same as the SD.
The mean of all trials is shown in the bottom of the table.
One immediately sees that 
$(\expec{V^2}/\expec{\sigma^2})^{1/2} <
\expec{\mach} < \expec{V^2/\sigma^2}^{1/2}$. 
If one were to assume a Gaussian distribution for $\mach$,
the standard deviation of the mean is ${\rm SD}/\sqrt{N} \ltsim 0.04$,
where, for the $R=10$ and 20$\himpc$ cases, $N$ is the number of 
independent spheres which fit in the simulation box.
However, we will show in the next section that, 
for $R=5$ and 10$\himpc$, 
the distribution of $\mach$ is not well described by a Gaussian,
so ${\rm SD}/\sqrt{N}$ is not the correct error in these cases.

The trend in the simulated Mach number is as follows:
$\mach_{dm-pt}<\mach_{gal-pt}<\mach_{gal-gp}<\mach_{dm-gp}$,
where the lower indexes indicate the different methods of calculation 
as explained in \S~\ref{method_section}. `dm-pt' refers
to both `galctr' and `dmctr' cases of the particle-based DM calculations. 
For the particle-based calculations, the Mach numbers 
using different centers and velocity tracers tend to 
converge one another on large scales.
In the group-based calculation, the difference between 
$\mach_{gal}$ and $\mach_{DM}$ is apparent. 
We have confirmed that the same trend is observed in our new 
$\lbox=25\himpc$ simulation as well when the same calculation
was performed with a 5$\himpc$ tophat patch.

To understand where these differences in $\mach$ arise, 
we summarize the mean and the rms value of $V$ and $\sigma$
in Table~\ref{table2} and \ref{table3}.
From these two tables and the same calculation with the $\lbox=25\himpc$ 
on $R=5\himpc$, the robust trends we see on scales $R\gtsim 5\himpc$ are the following:
$V_{gal-gp}<V_{particle-based}<V_{dm-gp}$
and
$\sigma_{group-based}<\sigma_{gal-pt}<\sigma_{dm-galctr-pt}<\sigma_{dm-dmctr-pt}$.
Differences within $\sim 20\kms$ are statistically insignificant,
but there are some cases that the difference amounts to $\sim 60\kms$, 
although still within one standard deviation.  
We have also carried out the same calculations on the $R=1\himpc$ scale, 
and find that the first inequality of the bulk flow shown above does not hold
in both $\lbox=25$ and 100$\himpc$ simulation. 
Also, for all cases of $\lbox=100\himpc$, we find $\sigma_{dm-gp}<\sigma_{gal-gp}$,
but the opposite relation is found in our new $\lbox=25\himpc$ on the scale
of $R=5\himpc$.

\notetoeditor{Place Table~\ref{table2} and \ref{table3} here.}

The difference in the bulk flow between the
particle-based and group-based calculation can be ascribed to 
the way it was calculated. In the particle-based calculation, 
we weighted each particle velocity by its particle mass, but in
the cases of the group-based calculations, we put equal weight on each 
galaxy or DM halo to mimic the real observation. We confirmed that, 
if we weight by the object's mass in the group-based calculation, 
the bulk flows reduced to the same values as the mass-weighted 
particle-based calculations. 

For the velocity dispersion, it is natural to see that the 
$\sigma_{group-based}<\sigma_{particle-based}$, 
as the internal velocity dispersion is erased by the grouping. 
Also, we expect to see $\sigma_{gal-pt}<\sigma_{dm-pt}$, 
as galaxy particles have formed out of sticky gaseous material 
compared to collisionless DM particles.

So, we regard the following relations as the most robust trends observed 
in our simulations:
1) $V_{dm-gp}>V_{gal-gp}$, and $V_{dm-gp}$ is always larger than any other cases
 (only for non-mass-weighted calculations); 
2) $\sigma_{group-based}<\sigma_{particle-based}$;
3) $\mach_{group-based}>\mach_{particle-based}$.


To summarize, our calculations show that the different methods of 
calculation result in different values of bulk flow and 
velocity dispersion, hence different Mach numbers as well.
We find that the grouping affects the resulting Mach number.
However, the differences in the simulated $\mach$ are smaller than 
the discrepancies between the simulated and the observed $\mach$, 
so they are not significant enough to change the arguments to follow.

\subsection{Distribution of $V$, $\sigma$, and $\mach$}
\label{distribution_section}
One would like to understand how the observed $\mach$ 
compares with the distribution of the simulated $\mach$,
and how it arises from the distribution of $V$ and $\sigma$. 

\begin{figure}
\plotone{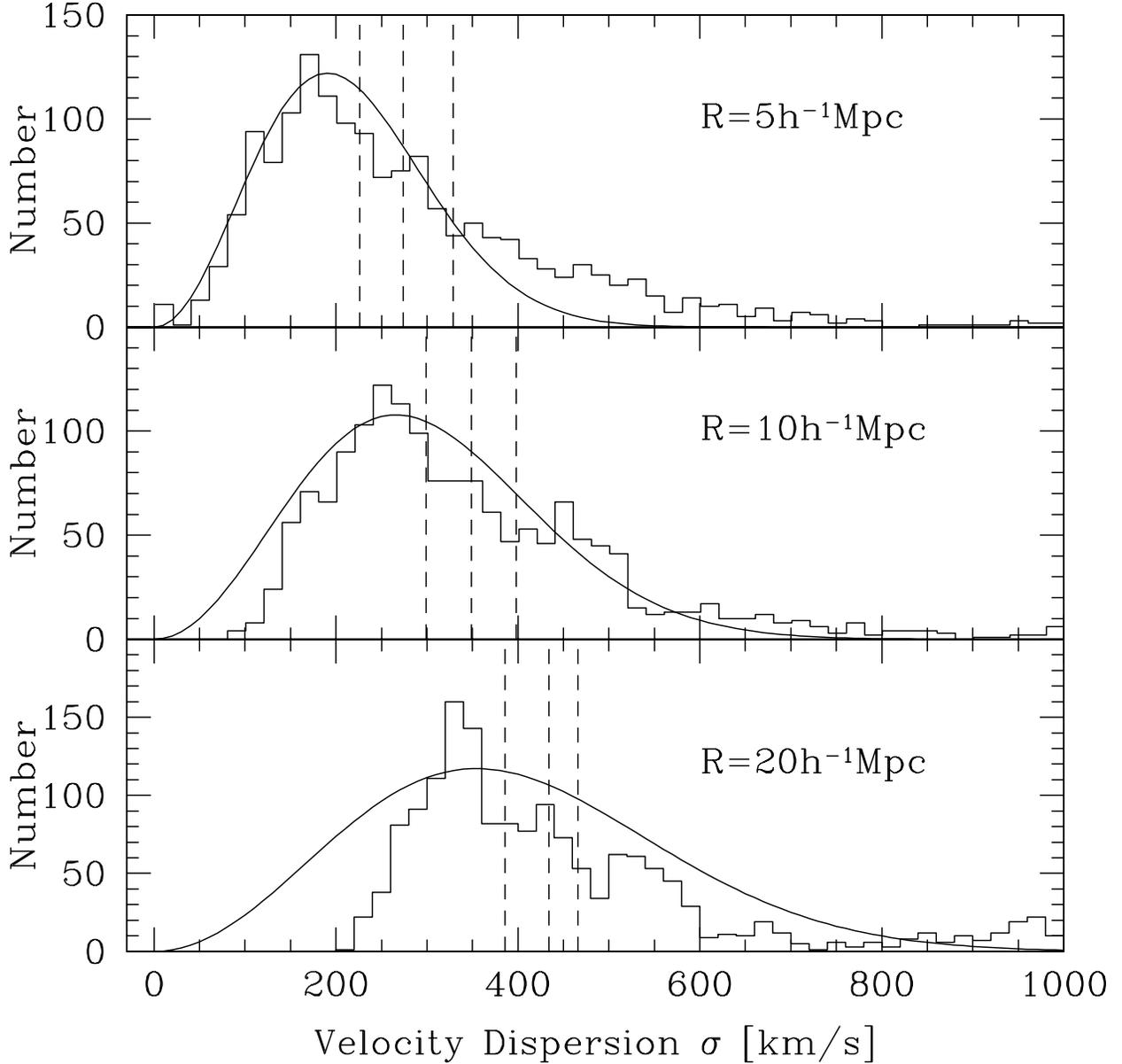}
\caption{
Distribution of the velocity dispersion of the simulated
galaxies (`gal-gp' case) for $R=5$, 10, and 20$\himpc$ 
from top to bottom.
The smooth solid curves show the `eyeball' fits to Maxwellian distribution.
The three vertical dotted lines in each panel are, from left to 
right, the median, the mean, and the rms value of the
distribution. For $R=5\himpc$, the simulated distribution has a longer tail than
Maxwellian. For $R=10$ and 20$\himpc$, the distribution have a sharper cutoff
at low values, and is not well described by Maxwellian.
\label{disp_dist.ps}
}
\end{figure}

Theoretically, bulk flow is expected to follow a Maxwellian 
distribution. In Figure~\ref{bulk_dist.ps},
we have already shown that the simulated bulk flow 
can be described by a Maxwellian distribution fairly well.
The distribution of velocity dispersion is non-trivial.
In Figure~\ref{disp_dist.ps}, we show the distribution of the 
simulated $\sigma$ of the grouped galaxies (`gal-gp' case). 
The three vertical dashed-lines in each panel are the median,
the mean, and the rms values of the distribution.  
The solid curves are the `eye-ball' fits to a Maxwellian distribution.
For the $R=5\himpc$ case, it is fitted to a Maxwellian relatively well
except the longer tail at large values of $\sigma$. 
For $R=10$ and 20$\himpc$ case,
the distribution is not well characterized by Maxwellian.
The simulated distribution has a steeper cutoff at low values.

\notetoeditor{Place \Fig{disp_dist.ps} here.}
 
\begin{figure}
\plotone{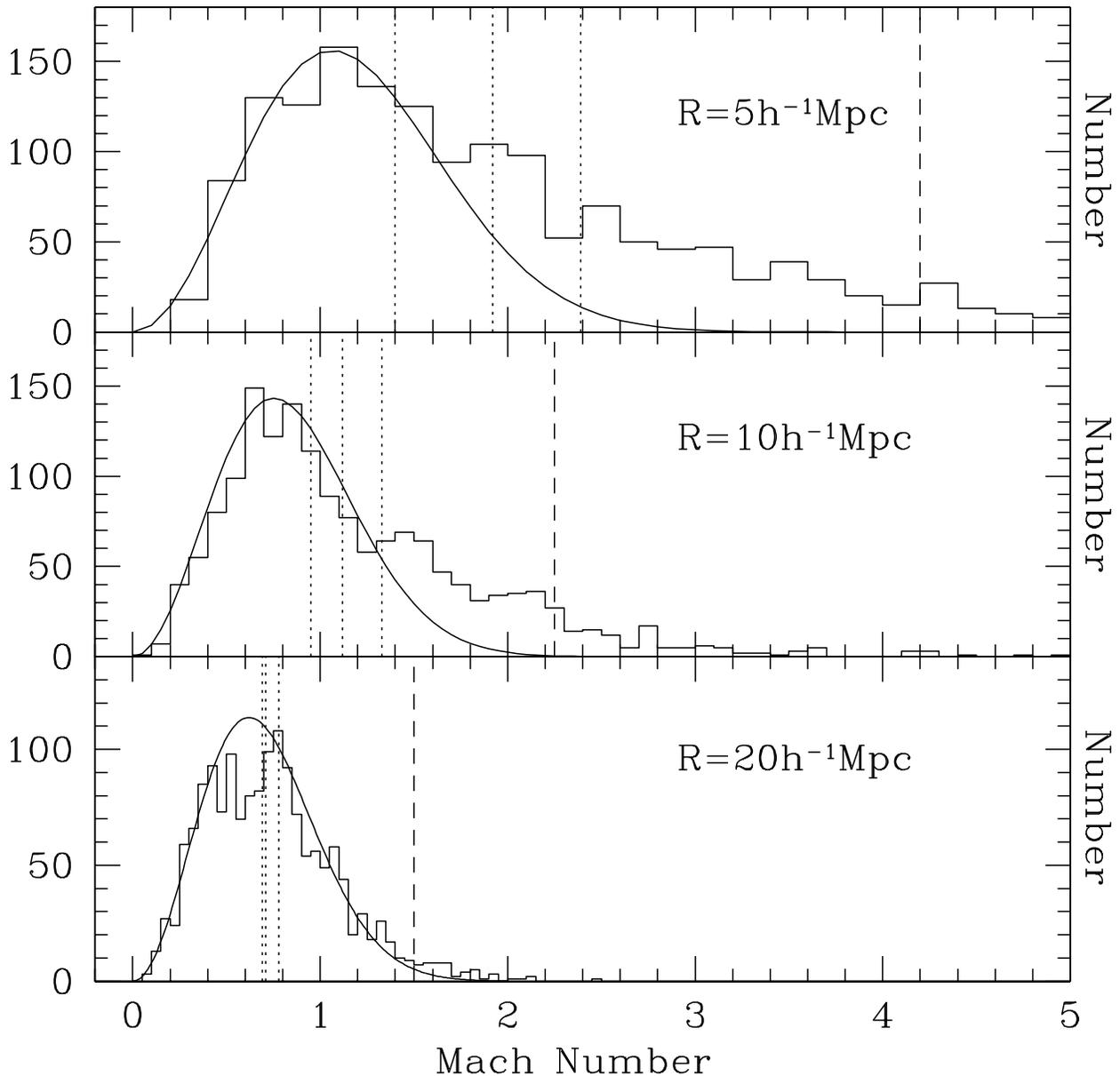}
\caption{
Mach number distribution of the simulated galaxies
(`gal-gp') for $R=5$, 10, and 20$\himpc$ from top to bottom.
This is after the bulk flow correction for the lack of long wavelength
perturbations. 
The smooth solid curves show the `eyeball' fits to Maxwellian distribution.
The three vertical dotted lines in each panel are, from left
to right, $(\expec{V^2}/\expec{\sigma^2})^{1/2}$, $\expec{\mach}$, and
$\expec{V^2/\sigma^2}$ as summarized in Table~\ref{table1}.
The observed $\mach$ (dashed line) is higher than the simulated mean $\expec{\mach}$
by more than 2-standard deviations at 92, 94, and 71\% confidence
level for $R=5, 10$, and 20$\himpc$ cases, respectively. 
See \S~\ref{distribution_section} for discussion.
\label{mach_dist.ps}
}
\end{figure}

In \Fig{mach_dist.ps}, we show the distribution of the simulated
$\mach$ of the grouped galaxies (`gal-gp' case). 
The smooth solid curves show the `eyeball' fits to a Maxwellian 
distribution. For the $R=5$ and $10\himpc$ cases,
the simulated Mach number distribution has a longer tail than
does the Maxwellian distribution. At the scale of $R=20\himpc$, 
the distribution is well fitted by a Maxwellian distribution.
For all the other methods of the calculation listed in Table~\ref{table1}, 
we find the same qualitative behavior.
We note that \citet{Suto90} have argued that the Mach number is 
distributed slightly broader than Maxwellian, consistent with our result.
The three vertical dotted lines in each panel are, from left
to right, $(\expec{V^2}/\expec{\sigma^2})^{1/2}$, $\expec{\mach}$, 
and $\expec{V^2/\sigma^2}$ as summarized in Table~\ref{table1}. 
Because of the long tail in the distribution for the $R=5$ and $10\himpc$
cases, the rms Mach number and the simple mean $\expec{\mach}$
do not reflect the peak of the distribution well. 
The dashed lines on the right show the observed $\mach$,
which will be summarized in the next section. 
The observed $\mach$ is higher than the mean $\expec{\mach}$ by 
more than 2-standard deviations at the 92, 94, and 71\% confidence level 
for $R=5, 10$, and 20$\himpc$ cases, respectively.

\notetoeditor{Place \Fig{mach_dist.ps} here.}

\begin{figure}
\epsscale{0.40}
\plotone{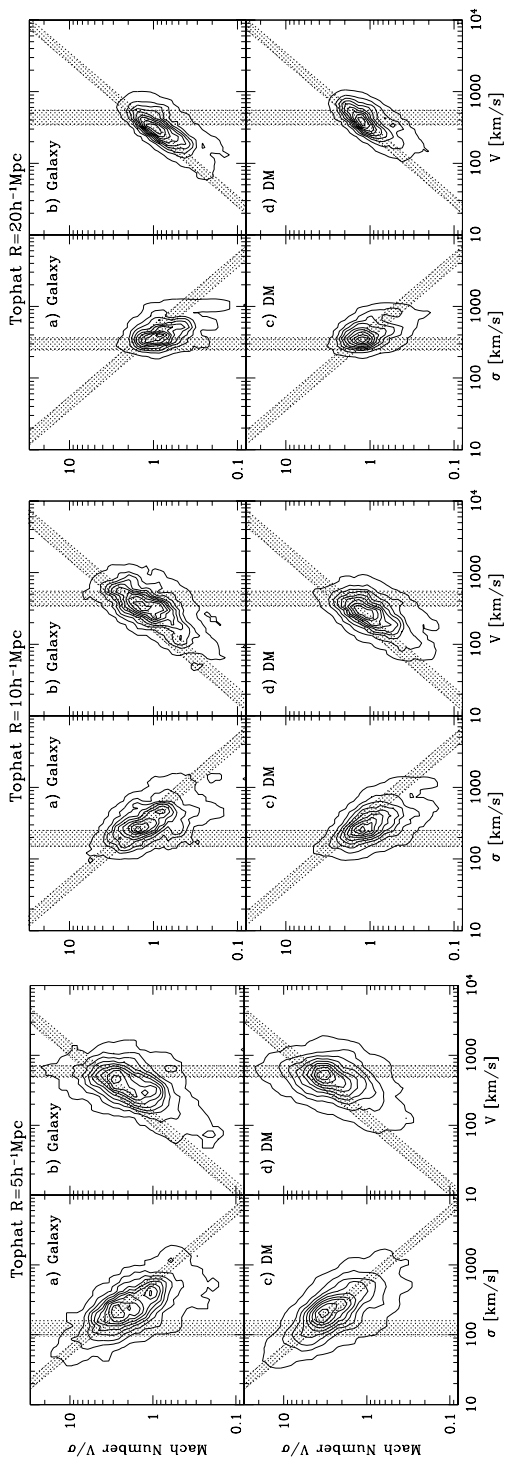}
\caption{
Number density distribution of the tophat patches in  
$\mach - \sigma$ and $\mach - V$ plane for the 
patch sizes of $R=5, 10$ and $20\himpc$ (group-based calculations).
Contours are equally spaced in logarithmic scale. 
The grey strips indicate the
`best-guess' velocity range from the observations summarized 
at the end of \S~\ref{overdensity_section}. 
See \S~\ref{distribution_section} for discussion. 
\label{machall.ps}
}
\end{figure}

How does this unusually high-$\mach$ arise from the 
distribution of $V$ and $\sigma$?
In \Fig{machall.ps},  
we show the number density distribution of the simulated tophat patches on 
$\mach - \sigma$ and $\mach - V$ plane for the patch sizes of 
$R=5, 10$ and $20\himpc$ (group-based calculations). 
Contours are of the number-density distribution of the simulated sample 
on an equally spaced logarithmic scale. 
Overall, as the patch size $R$ increases, the bulk flow 
decreases and the velocity dispersion increases as we 
already saw in \Fig{bulk_dist.ps} and \ref{disp_dist.ps}.
This is what we naively expect in the Friedman universe:
$V$ is a monotonically decreasing function of $R$ approaching 
zero as the largest irregularities are smoothed over, and $\sigma$
grows monotonically, saturating at the scale where $V$ has leveled
off, at the same value that $V$ had on small scales (OS90).
One can also see that the distribution of $\mach$ shifts 
down as the patch size $R$ increases. This can be seen more clearly in 
\Fig{mach_scale_plain.ps} and \ref{mach_dist.ps}.

\notetoeditor{Place \Fig{machall.ps} here.}

The grey strips in \Fig{machall.ps} are the `best-guess' ranges 
of $V$ and $\sigma$ based on observations. 
We take $V=500-700\kms$ and $\sigma=100-160\kms$ for $R=5\himpc$, 
$V=350-550\kms$ and $\sigma=150-250\kms$ for $R=10\himpc$, 
and $V=350-550\kms$ and $\sigma=250-350\kms$ for $R=20\himpc$ 
based on the observed range of values summarized in 
the end of next section. These ranges correspond to
$\mach=4.6\pm 1.3$ ($R=5\himpc$), $2.3\pm 0.8$ ($R=10\himpc$), and
$1.5\pm 0.4$ ($R=20\himpc$).
The tilted strips naturally arise from the definition
of the Mach number once we fix the value of either $V$ or $\sigma$;
$\mach \propto 1/\sigma$ or $\propto V$.
Note that the overlapping region of the two strips is off the peak 
of the entire distribution, as we already saw in Figure~\ref{mach_dist.ps}.
This offset is mainly caused by the observed low velocity dispersion.

\begin{figure}
\epsscale{1.0}
\plotone{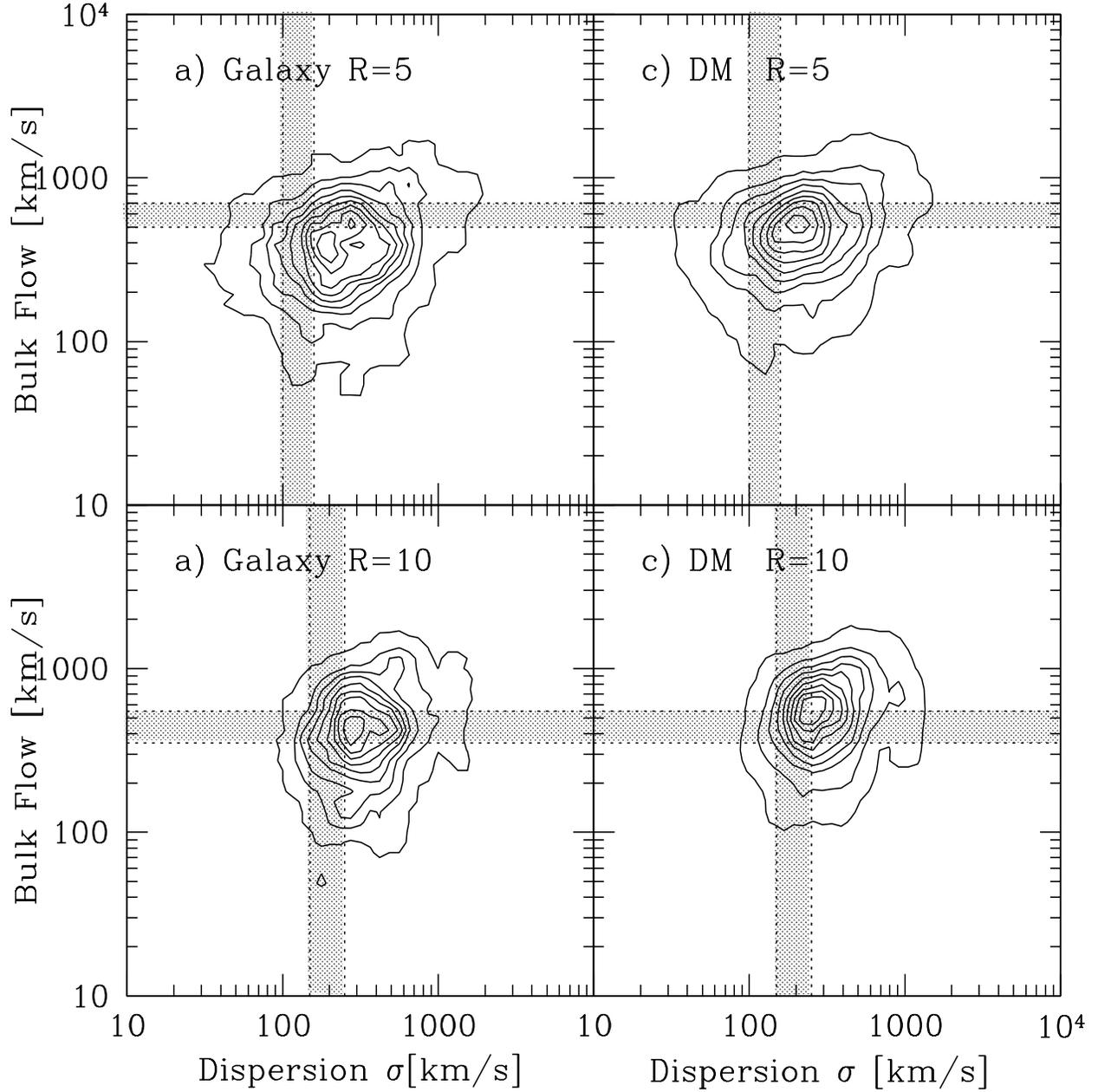}
\caption{
Velocity dispersion vs. bulk flow of the simulated galaxies 
(`gal-gp' case) for the $R=5$ and 10$\himpc$ cases.
There is a slight hint of positive correlation between the two quantities,
but otherwise they seem to be decoupled.
\label{disp_bulk.ps}
}
\end{figure}

In \Fig{machall.ps}, the abscissa and the ordinate are
not independent of each other because of the definition of the Mach number. 
To show the independent quantities on both axes, 
we show the bulk flow against 
the velocity dispersion of the simulated sample of grouped 
galaxies and DM halos for the case of $R=5$ and $10\himpc$ 
in Figure~\ref{disp_bulk.ps}. 
There is a slight hint of positive correlation between the two quantities,
but otherwise, they seem to be decoupled.
The grey strips are the same as in \Fig{machall.ps}.

\notetoeditor{Place \Fig{disp_bulk.ps} here.}

We further discuss the implication of the observed high Mach number of 
the Local Group in the next section by turning our eyes 
to the local overdensity.

\section{$V$, $\sigma$ and $\mach$ as a Function of Overdensity}
\label{overdensity_section}
In this section, we study the correlation between
$V$, $\sigma$, $\mach$ and local overdensity $\delta$.
We calculate $\delta$ at all sampling points
using spherical tophat patches of the same sizes as we used in 
calculating $V$, $\sigma$ and $\mach$.
For DM particles, we simply add the mass of all the particles in 
the patch and divide by the total mass in the simulation box to 
obtain the local mass-overdensity $\delta_{DM}$.
For galaxies, we use the updated isochrone synthesis model GISSEL99 
\citep[see][]{BC93} to obtain the absolute luminosity in V-band, 
and calculate the luminosity-overdensity $\delta_{L_V}$ in the 
same manner as the mass-overdensity. The GISSEL99 model takes 
the metallicity variation into account. Comparison of 
this simulation with various observations in terms of `light' is 
done by \citet{Nagamine00} in detail.
The use of luminosity-overdensity is not absolutely necessary here, 
and one should get the same conclusions as presented in this paper 
even if one uses the mass-overdensity of galaxies, since both 
overdensity roughly follow each other. 
We could in principle incorporate 
dust extinction by using a simple model, but that is a minor detail 
which would not change our conclusions in a qualitative manner.

\begin{figure}
\epsscale{0.40}
\plotone{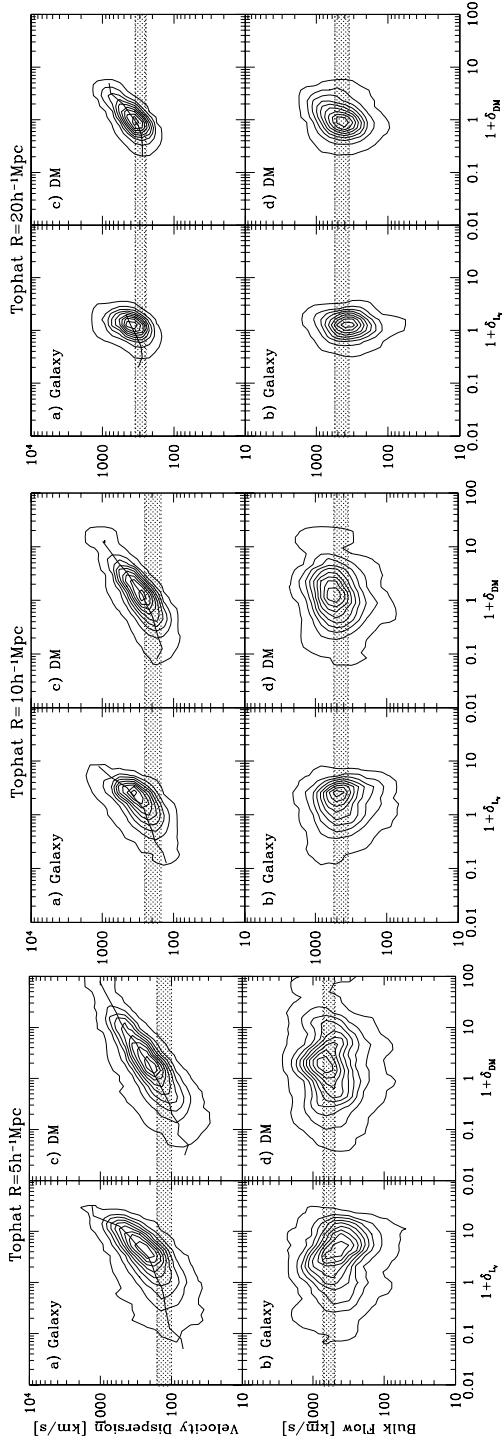}
\caption{
Velocity dispersion and bulk flow as a function of luminosity- and 
mass-overdensity for galaxies and DM halos, respectively, 
calculated with tophat $R=5, 10$ and $20\himpc$ patches 
(group-based calculation).
Contours are equally spaced in logarithmic scale. 
The grey strips indicate the `best-guess'
velocity range from the observations summarized at the end of 
\S~\ref{overdensity_section}. 
The solid line running through the velocity dispersion contour 
is the median within each bin of overdensity. 
See \S~\ref{overdensity_section} for discussion.
\label{vel_overdall.ps}
}
\end{figure}

In \Fig{vel_overdall.ps}, 
we show $V$ and $\sigma$ as functions of $\delta$ on scales of 
$R=5, 10$ and $20\himpc$, respectively (group-based calculation). 
The contour levels are the same as before. 
Again, the grey strips indicate the same `best-guess' range based on 
the observations as already described in the previous section. 
An important feature to note here is that $\sigma$ and $\delta$ are 
strongly correlated with each other, while $V$ and $\sigma$ are not. 
{\it Velocity dispersion is an increasing function of overdensity}.
This correlation between $\delta$ and $\sigma$ is 
similar to that seen in the case of $\sigma_{12}$, as described 
in \S~\ref{introduction_section}.
In the case of $\sigma$, it is weighted by the pairs always taken 
relative to the center-of-mass velocity (bulk flow $V$) of the patch, 
whereas in the case of $\sigma_{12}$, one takes all possible 
pairs in the patch. 
The solid line running through the contour in $\sigma - \delta$ plot 
indicates the median of the sample in each bin of overdensity. 
\citet{Willick98} studied the small-scale velocity dispersion 
in the observed data under the assumption of a linear relation between 
$\sigma$ and $\delta$, but our calculation predicts a shallower 
power-law dependence of $\sigma \propto \delta^{0.3-0.5}$ on all scales,
with the power index being larger at larger $\delta$.

\notetoeditor{Place \Fig{vel_overdall.ps} here.}

\begin{figure}
\epsscale{0.4}
\plotone{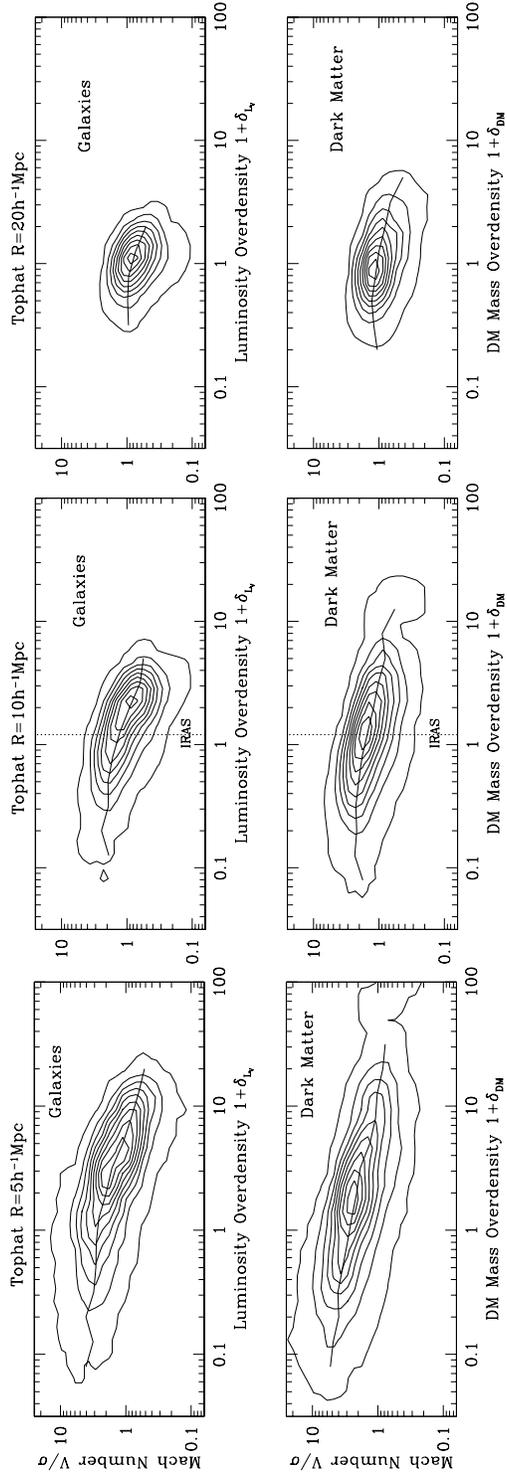}
\caption{
Cosmic Mach number as a function of luminosity- and mass-overdensity for 
galaxies and DM halos, respectively, calculated with tophat $R=5, 10$ and
$20\himpc$ patches (group-based calculation).
Contours are equally spaced number density distribution of the 
simulated sample in logarithmic scale. 
The solid line is the median within each bin of overdensity.
{\it The cosmic Mach number is a weakly decreasing function of local
overdensity}.
See \S~\ref{overdensity_section} for discussion.
\label{mach_overdall.ps}
}
\end{figure}

We then plot $\mach$ against $\delta$ in \Fig{mach_overdall.ps} 
on scales of $R=5, 10$ and $20\himpc$ (group-based calculation). 
The contour levels are the same as before. 
{\em The cosmic Mach number is a weakly decreasing function of overdensity}. 
This correlation between $\mach$ and $\delta$ originates from 
that between $\delta$ and $\sigma$.
Roughly speaking, low overdensity suggests low $\sigma$ and large $\mach$.
Therefore, the observed high-$\mach$ of the Local Group compared to the 
mean suggests that the Local Group is likely to be located in a 
relatively low overdensity region if our model is correct.
We note that \citet{Weygaert99} reach a similar conclusion by 
simulating the Local Group using constrained initial conditions.
However, it is also important to note that a given $\mach$ does not correspond
to a single value of $\delta$ due to both the weakness of the correlation 
and the significant scatter around the median which is indicated 
by the solid line.

\notetoeditor{Place \Fig{mach_overdall.ps} here.}

The dotted vertical line in the $R=10\himpc$ panel in \Fig{mach_overdall.ps} 
indicates $1+\delta_{IRAS}=1.2$ 
which is the observed IRAS galaxy number-overdensity at the Local Group 
\citep{Strauss95}
(It was calculated with a $R=5\himpc$ Gaussian window
which corresponds to $R=5\sqrt{5}=11.2\himpc$ tophat window).
It shows that the Local Group is off the peak of the distribution 
for galaxies, supporting our statement.
The fact that the IRAS survey samples only star forming galaxies
which tend to reside in low-density regions is not so important here
since it is only an issue in the centers of clusters.

\begin{figure}
\epsscale{1.}
\plotone{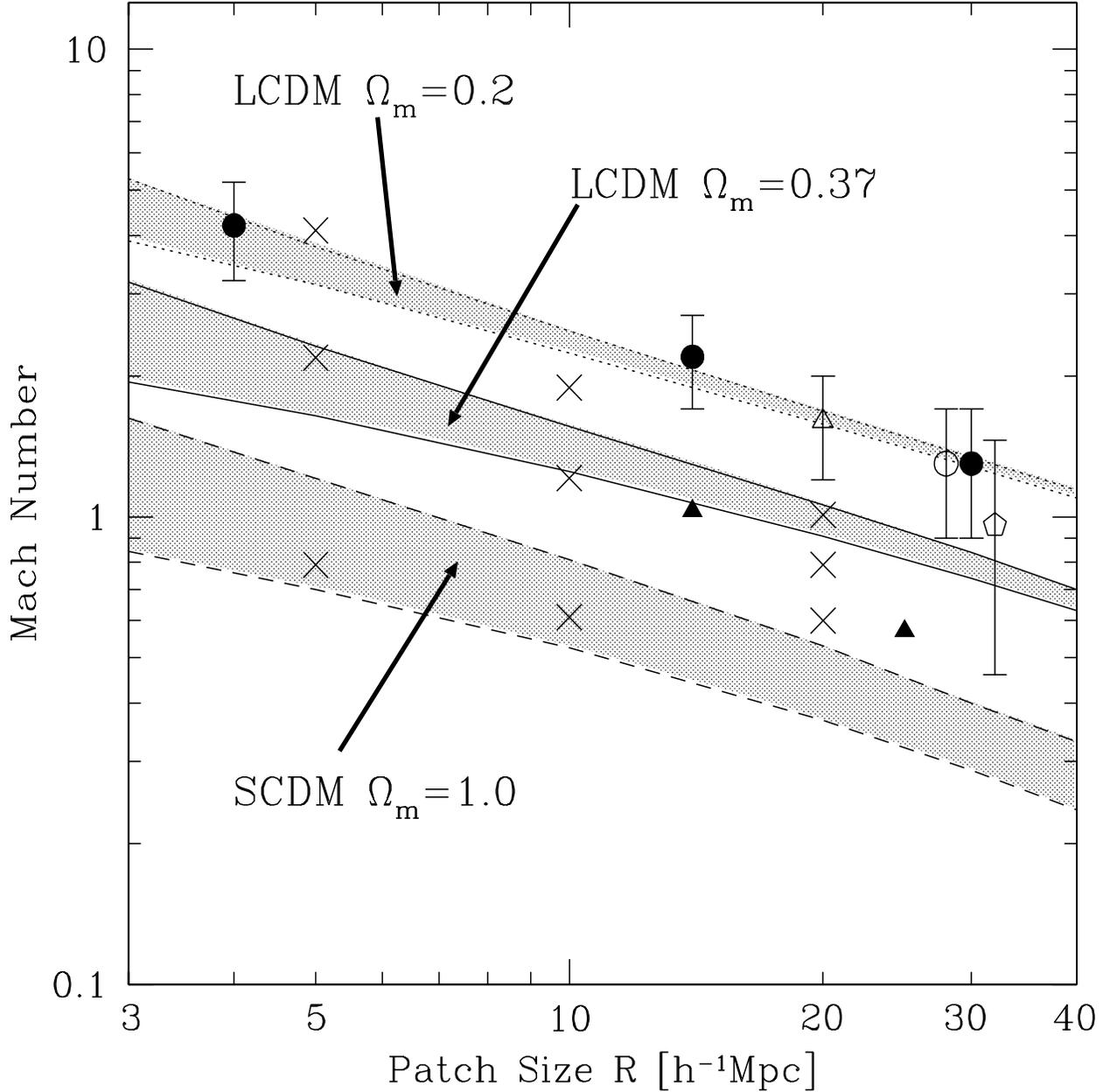}
\caption{
Cosmic Mach number as a function of scale $R$. The three crosses 
at each scale are the mean Mach number of the simulated galaxies 
divided in terms of the local overdensity:
the top is the 1st quartile (in highest density regions), 
the bottom is the 4th quartile (in lowest density regions), 
and the middle is the mean of the total sample.
The three shaded regions are the linear theory predictions 
of the rms Mach number similar to those in Figure~\ref{mach_scale_plain.ps},
where the top boundaries are calculated with the full COSMICS power spectra
using the mass density indicated in the figure, and the bottom
boundaries are calculated with those which were evolved to non-linear
regimes by \cite{Peacock96} scheme. 
The observed data points are summarized in the end of 
\S~\ref{overdensity_section}. 
Note that the lowest density quartile has a larger 
mean Mach number than do the highest density quartile. 
Many of the observational estimates appear
to be consistent with the $\Omega_m=0.2$ line than the 
simulated value of $\Omega_m=0.37$. 
See \S~\ref{overdensity_section} for discussions.
\label{mach_scale_overd.ps}
}
\end{figure}

To illustrate the above point more clearly, we divide the simulated galaxy sample 
into quartiles of local overdensity, and calculate $\expec{\mach}$ 
for each quartile. In \Fig{mach_scale_overd.ps}, the three crosses 
at each scale are the mean of each quartile of the grouped galaxies: 
top (1st quartile; lowest-$\delta$), bottom (4th-quartile; 
highest-$\delta$), and middle (total sample). 
Note that the galaxies in low-density regions have higher $\mach$.
We will discuss this correlation further in the next section in relation 
to the galaxy age. 
The solid, dotted, and dashed-lines are the linear theory predictions
calculated with the full linear COSMICS $P(k)$ for the indicated $\Omega_m$, 
similar to those in Figure~\ref{mach_scale_plain.ps}.

Let us now turn to the observations which are shown in the figure as well.
The three solid circles in \Fig{mach_scale_overd.ps} are 
the observational 
estimates made by OS90: $\mach(R=4\himpc) = 4.2\pm 1.0$ 
from $V=550\pm40\kms$ \citep{Lubin85} and $\sigma=130\pm30\kms$ 
\citep{Rivolo81, Sandage86, Brown87}; 
$\mach(R=14\himpc) = 2.2\pm0.5$ from $V=450\pm90\kms$
and $\sigma=205\pm10\kms$ \citep{Groth89};
$\mach(R=30\himpc) = 1.3\pm0.4$ from $V=500\pm130\kms$
and $\sigma=375\pm30\kms$ \citep{Groth89}.
The two solid triangles are the estimates made by S93,
but note that they have adopted a modified definition of $\mach$: 
${\tilde \mach}(R=14\himpc)=1.03$
from 206 galaxies in the infrared Tully-Fisher (TF) spiral galaxy catalog of 
\citet{Aaronson82}; ${\tilde \mach}(R=25\himpc)=0.57$ 
from 385 galaxies in the $D_n-\sigma$ elliptical galaxy catalog of \citet{Faber89}. 
A more recent sample is the surface brightness fluctuation (SBF)
survey of 300 elliptical galaxies by \citet{Tonry00}. They find
$V\simeq300\pm150\kms$ and $\sigma=312\pm24\kms$ at the scale of $R=30\himpc$,
which yields $\mach_{SBF}(R=30\himpc)=0.96\pm0.5$ (open pentagon). 
The recent survey of 500 TF-spiral galaxies by \citet{Tully00} finds 
$V\simeq400\pm100\kms$. 
Taking $\sigma=300\pm50\kms$ as a typical value, 
one obtains $\mach(R=30\himpc)=1.3\pm0.4$ (open circle), 
exactly same as the previous estimate by OS90 on the same scale.
The IRAS PSCz survey gives $V=475\pm75\kms$ \citep{Saunders00} 
using linear theory. Again, assuming $\sigma=300\pm50\kms$ yields 
$\mach(R=20\himpc)=1.6\pm0.4$ (open triangle). 
The Mach numbers from these new surveys seem to confirm that 
the observed $\mach$ is larger than the SCDM prediction, 
as originally pointed out by OS90. 
Bulk flows from other surveys on scales larger than $R=30\himpc$ 
are summarized in \citet{Dekel00}. 

Although we made new estimates of the Mach 
number on scales $R\gtsim20\himpc$, these numbers should be 
regarded as tentative since the observed bulk flow on large scales 
still seems uncertain in the literature 
\citep[see][]{Courteau00, Dekel00}. 
But if these estimates are correct, we consider that the high
observed $\mach$ reflects the fact that the Local Group is located
in a relatively low-density region as we argued earlier in this section.

Another possibility to resolve the discrepancy between the simulated
$\expec{\mach}$ and the observed $\mach$ is that the real Universe has 
a lower mass density than the simulated value of $\Omega_m=0.37$. 
We find in Figure~\ref{mach_scale_overd.ps} that $\Omega_m=0.2$
line fits all the observational estimates very well.  
If indeed $\Omega_m=0.2$, the observed low velocity dispersion of galaxies and 
the high Mach number would be typical in such universes.

\notetoeditor{Place \Fig{mach_scale_overd.ps} here.}

One might wonder if our result would be significantly altered
were the power spectrum to be steepened by one of the various
mechanisms being proposed to solve the putative problems of the
CDM paradigm on small scales \citep[\eg,][]{Dalcanton00}. 
We explored one typical such variant, the warm dark
matter proposal, and found that for a particle mass in the permitted
range ($\gtsim 1{\rm keV}$, cf. \citealt{Narayanan00, Bode00}) the
effect on the expected Mach number is negligible because the 
turndown in the power spectrum occurs at such a high wavenumber
as to be unimportant on patch sizes greater than 1$\himpc$.

\section{Correlation between Galaxy Age, Overdensity and Mach Number}
\label{age_section}

\subsection{General Expectations}
Under the standard picture of hierarchical structure formation, larger
systems form from  mergers of small objects. Therefore, one naively 
expects that the DWARF galaxies which exist in the present 
day universe are the `left-overs' in the low-density regions, 
and the GIANTs to be located in high-density regions where DWARFs
gathered to form GIANTs. 
(We denote DWARFs and GIANTs in capital letters because we will 
symbolically divide our galaxy sample in the simulation 
into two sub-samples by their stellar mass.)
However, DWARFs which are about to merge into larger systems could 
also exist in high-density regions as well. 

\begin{figure}
\plotone{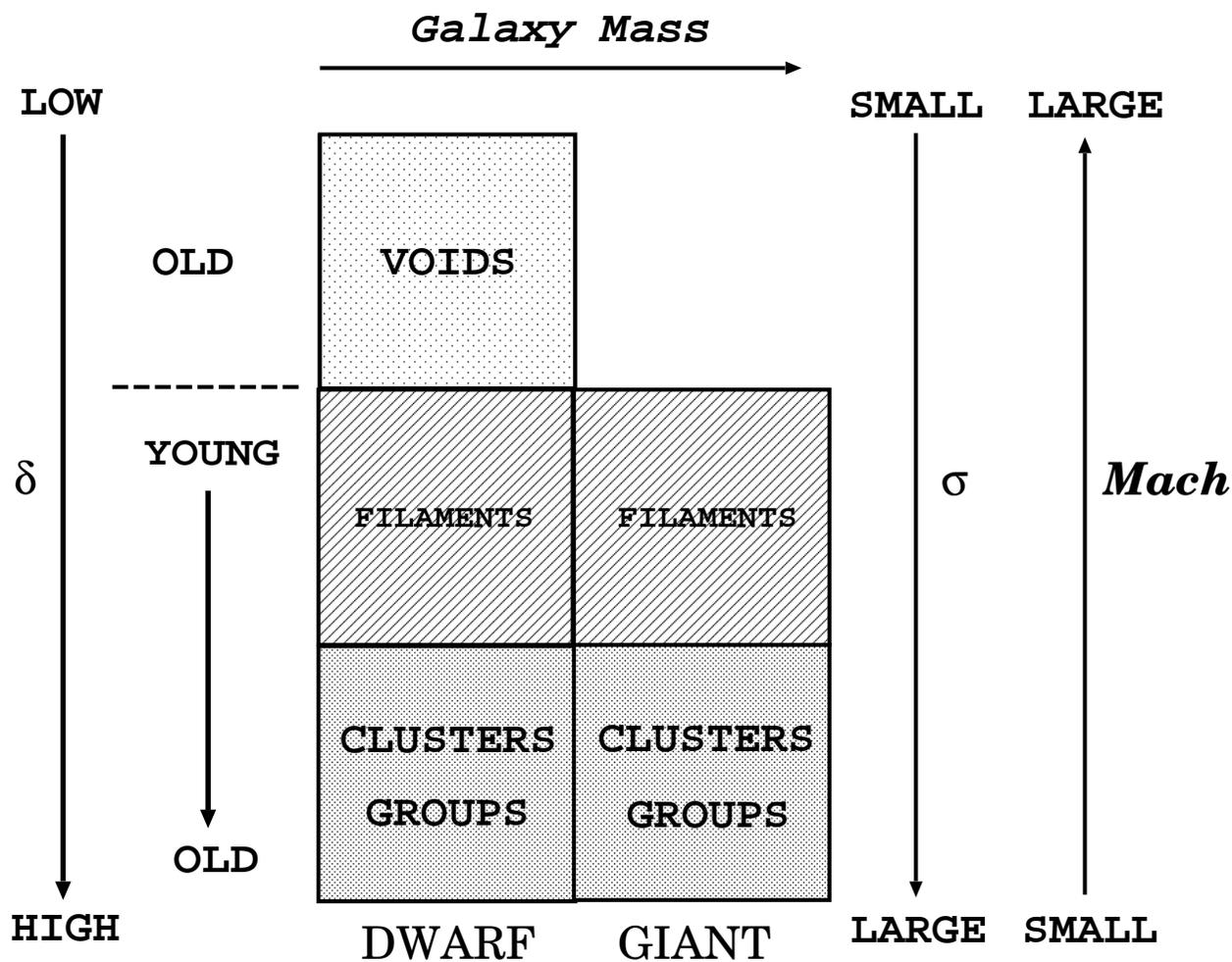}
\caption{
This figure summarizes the points made in this paper, and shows
various correlations. See text for discussion. 
The trends indicated by the arrows are confirmed by the simulation,
whose results are summarized in Table~\ref{table4} and \ref{table5}.
The left 3 boxes indicate the DWARF galaxies and the right 
2 boxes indicate the GIANTs.
The box is divided in terms of the local overdensity of the region
in which each population resides.
\label{dwarfgiant.ps}
}
\end{figure}

Now, let us define the formation time of a system in the simulation 
by the mass-weighted-mean of the formation time of the consisting 
galaxy particles.
Larger systems are the assembly of smaller systems which formed
earlier, so for GIANTs, the larger the system is, the older the formation 
time would be. 
(Note that we are using the terms `young' and `old' relative to the present, 
\ie, young $\equiv$ smaller $z_{form}$.)
DWARFs do not follow this trend, because some of the smallest DWARFs 
formed at very high-redshift will remain as it is without 
merging into larger systems.
Therefore, they are the oldest population by definition
despite the fact that they are the smallest systems.
This counter effect dilutes the correlation between age 
and local overdensity for the DWARF population.
Systems in high-density regions have larger 
$\sigma$, hence smaller $\mach$, and vice versa.
We summarize the above points in \Fig{dwarfgiant.ps}.
The three left boxes represent the DWARF galaxies divided
in terms of the local overdensity of the region they live in. 
DWARFs live in both low and high overdensity regions (VOIDS and CLUSTERS),
while GIANTs live in moderate to high overdensity regions (the right
two boxes). 
We denote the intermediate overdensity region as FILAMENTS.
The correlation with $\delta$, galaxy age, mass, $\sigma$ and $\mach$
is indicated by the arrows in the figure.

\notetoeditor{Place \Fig{dwarfgiant.ps} here.}

\subsection{Do we see the effect in the simulation?} 
\begin{figure}
\plotone{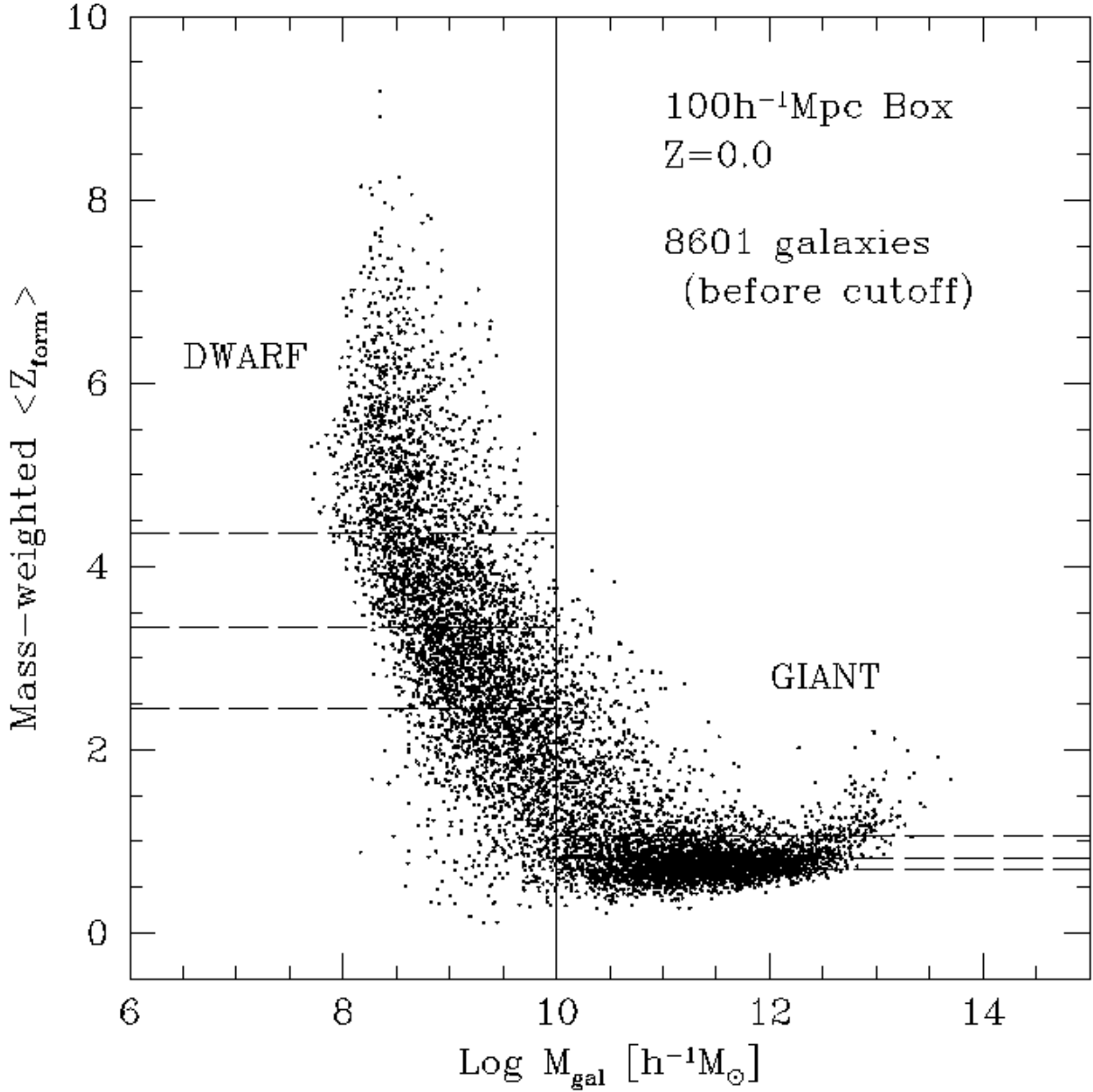}
\caption{
Mean formation time of the simulated galaxies at $z=0$ 
(converted to redshift) vs. stellar mass of galaxies. 
Mean formation time of each galaxy was calculated by taking 
the mass-weighted average of the formation time of 
consisting galaxy particles. All galaxies in the simulation box are shown.
The vertical line at $M_{gal}=10^{10}\hinv\Msun$ divides the sample
into DWARFs and GIANTs. The horizontal dashed lines are the 
boundaries of the quartiles in terms of $z_{form}$. See \S~\ref{age_section}
for discussions. 
\label{age_mass.ps}
}
\end{figure}

To see the above effect in the simulation, we divide the simulated 
galaxy sample in the simulation into DWARFs and GIANTs 
at the median mass of $M_{galaxy}=10^{10} \hinv M_{\odot}$
as shown in Figure~\ref{age_mass.ps}. 
Note that the galaxies shown in this figure was taken 
from $z=0$ output of the simulation, therefore, GIANTs
that formed at late times certainly include galaxy particles
that formed very early on.
We also divide each sample into quartiles by their
formation time. The formation time of each galaxy is calculated
as defined above, and is converted to redshift ($z_{form}$). 
The boundary redshift of the quartiles are 
shown as the horizontal dashed lines in Figure~\ref{age_mass.ps},
which are $z_{boundary}=2.44, 3.33, 4.36$ for DWARFs, 
and 0.68, 082, and 1.05 for GIANTs.
One sees from this figure that the trend is not as clear-cut as 
we naively expected above, though the basic line was correct. 
The formation time of DWARFs ranges widely, and the heavier DWARFs 
tend to be younger.  Very small DWARFs form at very high redshift 
($z_{form}\gtsim 4$), and the moderate size DWARFs continues to 
form through $z \sim 1$. GIANTs mainly form at moderate 
redshifts ($1<z_{form}<2$) when the global star formation rate is 
most active in the simulation \citep{Nagamine00}. 
One sees that the very massive GIANTs have a slight positive slope 
as we expected above. But near the boundary of DWARF and GIANT,
there are some less massive GIANTs that are older than the heavier GIANTs 
as well. 

\notetoeditor{Place \Fig{age_mass.ps} here.}

\begin{figure}
\plotone{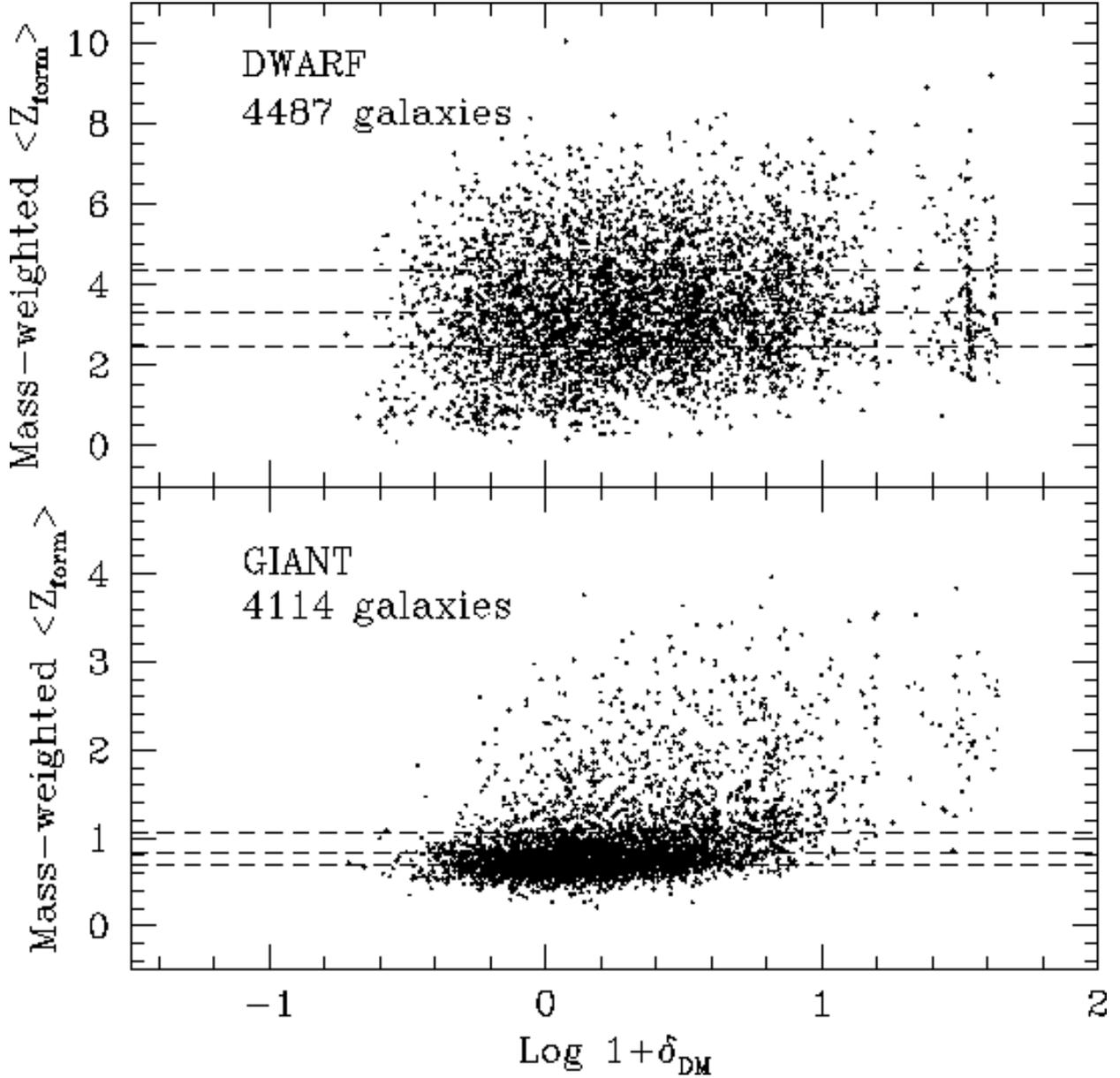}
\caption{
Mean formation time of the simulated galaxies (converted to redshift) 
vs. local overdensity (calculated with a tophat $R=5\himpc$ window).
Mean formation time of each galaxy was calculated by taking the 
mass-weighted average of the formation time of consisting galaxy particles. 
The sample is divided into DWARFs (top panel) and GIANTs (bottom panel) 
at $M_{gal}=10^{10}\hinv\Msun$ (stellar mass) as shown in Figure~\ref{age_mass.ps}.
The three horizontal dashed-lines in each panel are the boundaries
of the quartiles of the sample, divided in terms of formation time. 
DWARFs reside in all environments with a weak positive correlation 
between formation redshift and overdensity, while the GIANTs in very high-density 
regions tend to be older (larger $z_{form}$). See \S~\ref{age_section}
for discussions. 
\label{age_overd.ps}
}
\end{figure}

Now we discuss the correlation between overdensity,
galaxy age, and the Mach number.
In \Fig{age_overd.ps}, we show the formation time of galaxies 
as a function of DM mass-overdensity $\delta_{DM}$ calculated
with tophat patch of $R=5\himpc$.
One sees that DWARFs exist in all environments with a weak
positive correlation between $z_{form}$ and $\delta_{DM}$,
and that some older (\ie, larger $z_{form}$) GIANTs tend to 
be in high-density regions than less massive ones as suggested 
in \Fig{dwarfgiant.ps}.
The horizontal dashed lines indicate the boundaries of the 
quartiles in mean galaxy age.
In Table~\ref{table4}, we summarize the mean overdensity
of each quartile calculated with a $R=5\himpc$ tophat window. 
It is apparent from the table that the older systems reside in
higher density regions.
The contrast is less dramatic for the DWARFs than the GIANTs 
because the correlation between age and overdensity for DWARFs
is diluted by the old DWARFs located in low-density regions.

\notetoeditor{Place \Fig{age_overd.ps} and Table~\ref{table4} here.}

\begin{figure}
\plotone{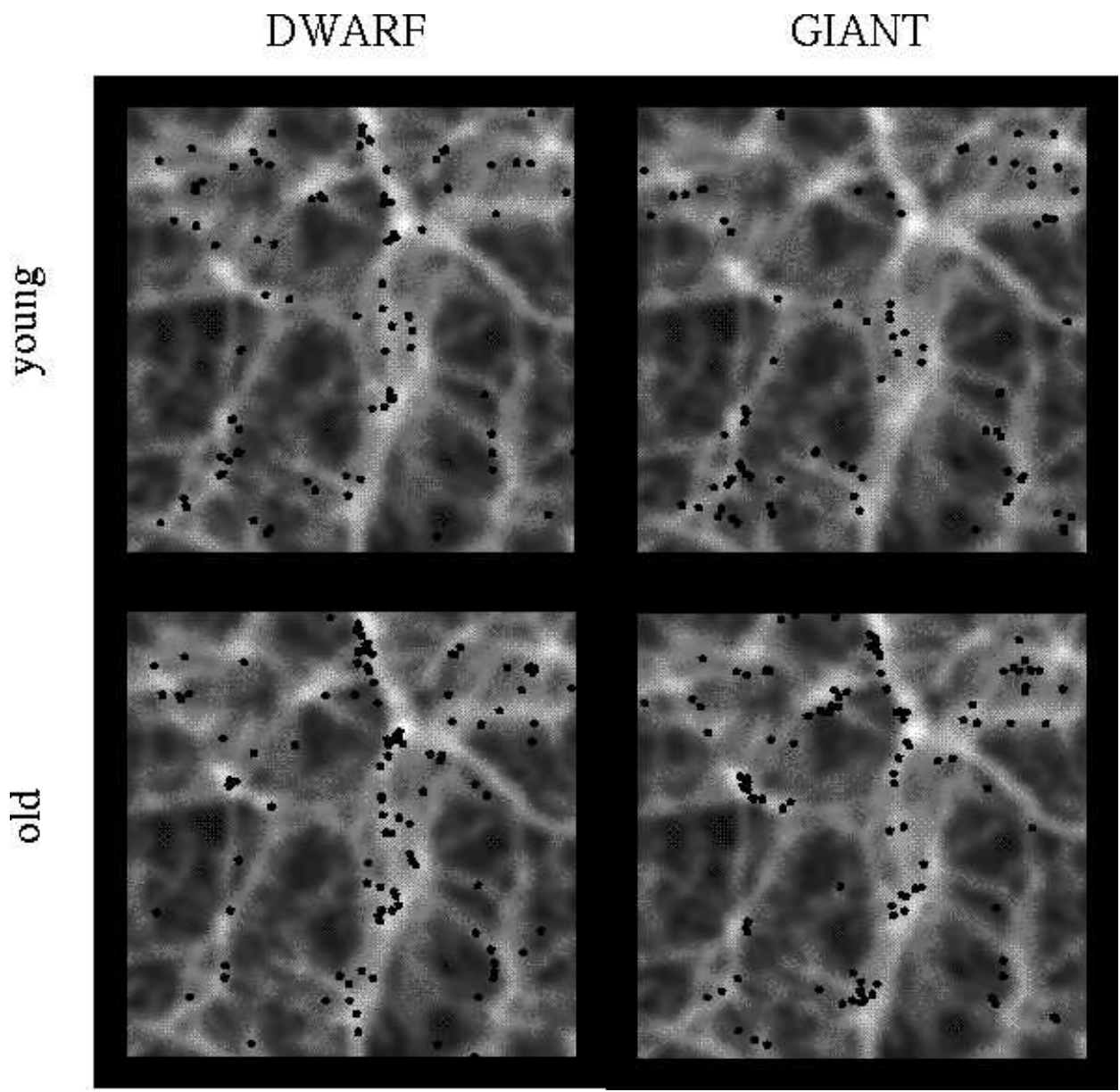}
\caption{
Projected slice of $5\himpc$ thickness from the simulation.
$\lbox=100\himpc$ on each side of the box. 
The smoothed DM density field is shown in the background and the 
location of each species of galaxies is shown with the solid points. 
Older population are more clustered than younger population.
Some old DWARF galaxies reside in low-density regions as well. 
GIANTs are well clustered in high-density regions with low Mach 
number. See \Fig{corr.ps} for correlation function, and 
Table~\ref{table5} for the mean Mach number of different samples.
\label{slice.ps}
}
\end{figure}

As a visual aid, we show a slice of $5\himpc$ thickness from the 
simulation in \Fig{slice.ps}. The smoothed DM density
field is in the background and the location of the galaxies 
is indicated by the solid points. One can clearly see that the older 
population is more clustered than the younger population for both 
DWARFs and GIANTs. Some old DWARF galaxies reside in low-density 
regions as well. GIANTs are more clustered in high-density regions.  
But it is a little difficult to see the difference between the old DWARFs 
and old GIANTs, or young DWARFs and young GIANTs.
Notice also the projection effect that some galaxies appear just by the 
filaments.

\notetoeditor{Place \Fig{slice.ps} here.}

\begin{figure}
\epsscale{0.5}
\plotone{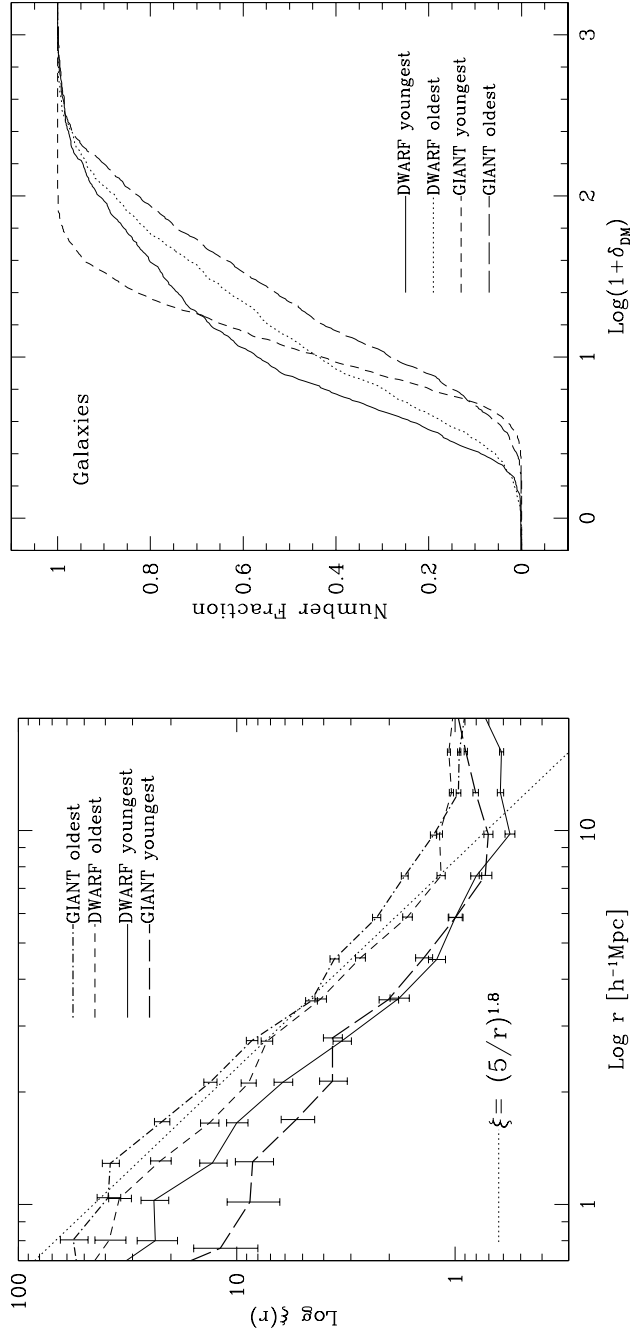}
\caption{
Left panel: Correlation functions of the oldest and the youngest
quartiles of DWARFs and GIANTs.
The oldest GIANTs are most strongly clustered, following the
well known power-law $\xi=(5/r)^{1.8}$.
The youngest galaxies are less clustered.
Right panel: Cumulative number fraction distributions of the 
oldest and the youngest quartiles of DWARF and GIANT galaxies
as functions of mass-overdensity (calculated with a tophat $R=1\himpc$ window).
Older galaxies tend to populate higher density regions
than younger galaxies. GIANTs prefer higher density regions
than DWARFs.
\label{corr.ps}
}
\end{figure}

To see the difference in the clustering property more clearly, 
we show the two-point correlation function 
of the oldest and the youngest quartiles of the DWARF and GIANT 
population in the left panel of \Fig{corr.ps}. 
The Poisson errorbars are shown together. 
As one expects, the oldest GIANTs are clustered most
strongly, and the oldest DWARFs are second strongly clustered,
but very close to the oldest GIANTs.
The two correlation functions of the oldest populations 
follow the power-law $\xi=(5/r)^{1.8}$ well on scales of
$1\himpc < R < 10\himpc$. 
The youngest DWARFs and GIANTs are less clustered,
and seems to be consistent with each other on 
scales of $3\himpc < R < 10\himpc$. 
The youngest GIANTs seem to have a weaker signal 
than the youngest DWARFs on scales less than $2\himpc$,
but it is not clear if this is real or a numerical artifact.

On the right panel of \Fig{corr.ps}, we show 
the cumulative number fraction distribution 
of different galaxy population as functions of 
local mass-overdensity (calculated with a tophat $R=1\himpc$ window). 
One sees that older galaxies tend to reside in 
higher density regions than younger galaxies, 
consistent with the correlation function shown 
in the left panel. GIANTs prefer higher density regions
than DWARFs.

\notetoeditor{Place \Fig{corr.ps} here.}

Finally, let us look at the correlation between overdensity and the 
Mach number. The mean Mach numbers of young and old galaxies 
are summarized in Table~\ref{table5}.
Here, `young' denotes the first 2 younger quartiles in galaxy age, 
and `old' denotes the 2 older quartiles.  
In all cases of GIANTs and $R=5\himpc$ of DWARFs, 
the older sample has smaller $\mach$ as expected. 
On larger scales ($R\gtsim5\himpc$), 
the patch starts to sample more DWARFs in low-density regions,
and the naively expected trend turns over in the opposite
direction for the DWARFs.

\notetoeditor{Place Table~\ref{table5} here.}

\section{Conclusions}
\label{conclusion}
We have studied the bulk flow, the velocity dispersion and 
the cosmic Mach number on scales of $R=5, 10$ and $20\himpc$
using a  LCDM hydrodynamical simulation, putting emphasis
on the environmental effects by the local overdensity,
and the correlation with galaxy age and size.
Different methods of calculation and the different definitions of 
$\mach$ were tried out to see the differences in the result.
We found $(\expec{V^2}/\expec{\sigma^2})^{1/2} <
\expec{\mach} < \expec{V^2/\sigma^2}^{1/2}$ (Table~\ref{table1}), 
and that the different methods of 
calculation result in different values of bulk flow and 
velocity dispersion, hence different Mach number as well.
We found that the grouping procedure affects the resulting 
Mach number significantly.
However, the difference in $\mach$ due to different methods of calculation 
is smaller than the discrepancy between the simulated and the observed $\mach$, 
therefore it is not significant
enough to change our following conclusions.  

We showed the distribution of the bulk flow, the velocity dispersion, 
and the Mach number in the simulation (Figure~\ref{bulk_dist.ps}, 
\ref{disp_dist.ps}, and \ref{mach_dist.ps}).
The bulk flows are fitted by a Maxwellian distribution well except that the 
uncorrected $R=5\himpc$ case has a longer tail. The velocity dispersion 
is not well fitted by a Maxwellian; it has a longer tail for $R=5\himpc$ case,
and a sharper cutoff at low values of $\sigma$ for $R=10$ and 20$\himpc$ cases.
As a result, Mach number is relatively well fitted by a Maxwellian, 
but with a longer tail for $R=5$ and 10$\himpc$ cases.

We discussed the theoretical predictions of
$\mach$ in \S~\ref{theory_section}, including the scale- and $\Omega_m$-dependence
of $\mach$. The range of the simulated Mach numbers fall just below the 
theoretical prediction (\Fig{mach_scale_plain.ps}), 
reflecting the non-linear evolution in the simulation 
which cannot be fully taken into account by simply plugging the non-linear 
power spectrum into the linear equations.
We also discussed in \S~\ref{theory_section} how we corrected the simulated 
bulk flow for the lack of long wavelength perturbations beyond 
the simulation box size.

The first main conclusion of this paper is that the observed velocity 
configuration of the Local Group is not the most typical one
if the adopted LCDM cosmology is correct.
Our calculation shows that the observed Mach numbers are 
higher than the simulated mean by more than 
2-standard deviations at high confidence levels (\Fig{mach_dist.ps}), 
and that the observed velocity configuration
is off the peak of the number density distribution 
in the $\mach-\sigma$ plane (\S~\ref{distribution_section}, \Fig{machall.ps}).
This discrepancy is mainly due to the low observed velocity dispersion, 
while the observed bulk flow is not that uncommon.

Second, we showed that {\it the cosmic Mach number is 
a weakly decreasing function of overdensity}
(\Fig{mach_overdall.ps}). 
The correlation originates from 
that between overdensity and the velocity 
dispersion (\Fig{vel_overdall.ps}).
This is a similar situation to that of the pairwise velocity dispersion. 
Roughly speaking, high Mach number suggests a low-density environment.
It is important to take this overdensity dependence of $\mach$ into account
in any analysis of cosmic Mach number or velocity dispersion, as is the 
case for the pairwise velocity dispersion.

Third, a few new observational estimates of $\mach$ were made 
in this paper on scales of $R=20$ and $30\himpc$ 
(\Fig{mach_scale_overd.ps}). 
They are much higher than the SCDM prediction, 
confirming the conclusions of earlier studies by other authors. 
Combined with our second point, 
the observed local high-$\mach$ is simply a reflection 
of the fact that the Local Group is in a relatively low overdensity region
as we know from the $IRAS$ survey.
Another possibility which resolves the discrepancy between the simulated
and the observed $\mach$ is that our Universe has a much lower
mass density than the simulated value of $\Omega_m=0.37$. 
If so, the observed low-$\sigma$ and high-$\mach$ would be 
typical in such universes.
As we showed in \Fig{mach_scale_overd.ps}, the observed Mach 
numbers are in good agreement with the linear theory prediction with
$\Omega_m=0.2$. This may be interpreted that the local value of 
$\Omega_m$ is closer to 0.2 instead of the simulated value of 0.37. 
We also explored the possibility of the warm dark matter proposal,
and found that for a particle mass in the permitted range 
($\gtsim 1{\rm keV}$, cf. \citealt{Narayanan00, Bode00})
the effect on the expected Mach number is negligible
on scales greater than $1\himpc$.

Fourth, 
we studied the correlation between galaxy mass, galaxy age,
local overdensity and the Mach number. 
Our major points are summarized in \Fig{dwarfgiant.ps}. 
Using the simulation, 
we showed that the older (redder) systems are strongly clustered 
in higher density regions with smaller $\mach$, 
while younger (bluer) systems tend to reside in lower density regions
with larger $\mach$ (\Fig{age_overd.ps}, \ref{slice.ps}, \ref{corr.ps} 
and Table~\ref{table5}),
as expected from the hierarchical structure formation scenario. 
We divided the galaxy sample into DWARFs and GIANTs 
in the simulation, and 
found that the GIANTs follow this expected trend, 
while DWARFs deviate from this trend on large scales ($R>5\himpc$) 
due to the presence of the old DWARFs in low-density regions
which did not merge into larger systems.
The two point correlation functions and the cumulative number fraction
distributions of different populations were presented.

\acknowledgments

 We thank Vijay Narayanan and Michael Strauss for useful discussions and comments.
We also thank Yasushi Suto for comments and providing us with 
a code to calculate the non-linear evolution of the power spectrum,
and Paul Bode for helping us with the calculation of the warm dark matter 
power spectrum.
KN is partly supported by the Physics Department. 
This work was supported in part by grants AST~98-03137 and ASC~97-40300.

\appendix

\section{Galaxy Particle Formation Criteria in the Simulation}
\label{criteria}

The criteria for galaxy particle formation in each cell of the 
simulation are:
\begin{eqnarray}{}
1+\delta_{tot} &>& 5.5, \\
m_{\mathrm{gas}} &>& m_{\mathrm{J}} \equiv G^{-3/2} \rho_{\mathrm{b}}^{-1/2}
C_s^3 \left[1 +\frac{1+\delta_d}{1+\delta_b} 
\frac{\bar{\rho_d}}{\bar{\rho_b}}\right]^{-3/2}, \\
t_{\mathrm{cool}} &<& t_{\mathrm{dyn}} \equiv \sqrt{\frac{3\pi}{32
G\rho_{tot}}} \mathrm{,~and} \\
\nabla\cdot\vv{v} &<& 0
\end{eqnarray}
where the subscripts ``$b$'', ``$d$'' and ``$tot$'' refer to baryons,
collisionless dark matter, and the total mass, respectively.  
$C_s$ in the definition of the Jeans mass is the isothermal sound speed.
The cooling time is defined as $t_{cool}={\rm n_e k_B T}/{\rm \Lam}$, 
where ${\rm \Lam}$ is the cooling rate per unit volume in units of 
[${\rm ergs/sec/cm^3}$]. Other symbols have their usual meanings.



\clearpage

\setcounter{thetabs}{0}

\stepcounter{thetabs}
\begin{deluxetable}{lcccccccccccccccc}  
\tablenum{\tabnum}
\tablecolumns{17}  
\tablewidth{0pc} 
\rotate 
\tablecaption{Summary of the Cosmic Mach Number in the Simulation\label{table1}}  
\tablehead{
\colhead{} & 
\multicolumn{4}{c}{$R=5\himpc$}  & \colhead{} &
\multicolumn{4}{c}{$R=10\himpc$} & \colhead{} &
\multicolumn{4}{c}{$R=20\himpc$} & \colhead{} & \colhead{$N_{\rm patch}$}  \\  
\cline{2-5} \cline{7-10} \cline{12-15} \cline{17-17}
\colhead{} &  
\colhead{$(\frac{\expec{V^2}}{\expec{\sigma^2}})^{\frac{1}{2}}$} & 
\colhead{$\expec{\mach}$} & 
\colhead{$\expec{\frac{V^2}{\sigma^2}}^{\frac{1}{2}}$} & 
\colhead{SD} & \colhead{} & 
\colhead{$(\frac{\expec{V^2}}{\expec{\sigma^2}})^{\frac{1}{2}}$} & 
\colhead{$\expec{\mach}$} & 
\colhead{$\expec{\frac{V^2}{\sigma^2}}^{\frac{1}{2}}$} & 
\colhead{SD} & \colhead{} &
\colhead{$(\frac{\expec{V^2}}{\expec{\sigma^2}})^{\frac{1}{2}}$} & 
\colhead{$\expec{\mach}$} & 
\colhead{$\expec{\frac{V^2}{\sigma^2}}^{\frac{1}{2}}$} & 
\colhead{SD} & \colhead{} & \colhead{} }
\startdata 
\sidehead{Particle-based}
a) gal-pt  & 1.40 & 1.92 & 2.39 & 1.00 & & 0.95 & 1.12 & 1.33 & 0.45 & & 0.69 & 0.71 & 0.78 & 0.18 & & 1585 \\
b) dm-galctr-pt   & 1.14 & 1.46 & 1.72 & 0.64 & & 0.90 & 1.06 & 1.22 & 0.39 & & 0.71 & 0.74 & 0.82 & 0.17 & & 1585 \\
c) dm-dmctr-pt   & 1.16 & 1.76 & 2.67 & 0.88 & & 0.89 & 1.14 & 1.33 & 0.43 & & 0.69 & 0.76 & 0.84 & 0.18 & & 4142 \\
\tableline
\sidehead{Group-based}
a) gal-gp  & 1.43 & 2.07 & 2.77 & 1.29 & & 1.05 & 1.25 & 1.47 & 0.50 & & 0.76 & 0.80 & 0.88 & 0.19 & & 1585\tablenotemark{a} \\
b) dm-gp    & 1.72 & 2.69 & 3.62 & 1.69 & & 1.29 & 1.56 & 1.78 & 0.55 & & 1.00 & 1.06 & 1.14 & 0.22 & & 4142\tablenotemark{b} \\
\tableline	    	       		   	         	  	        
mean of all  & 1.48 & 2.36 & 3.11 & & & 1.06 & 1.29 & 1.52 & & & 0.77 & 0.81 & 0.89 & & & \\
\enddata  

\tablenotetext{a}{For $R=5\himpc$ case, $N_{{\mathrm patch}}=1574$.} 
\tablenotetext{b}{For $R=5\himpc$ case, $N_{{\mathrm patch}}=4124$.} 

\tablecomments{Mean and the rms value of the cosmic Mach number
is summarized. SD stands for standard deviation.
$N_{\rm patch}$ is the number of patches that were eligible in each analysis 
(we rejected those patches which contained only one galaxy). 
All numbers shown are after the multiplication by the factors of 
1.43 ($R=5\himpc$), 1.56 ($R=10\himpc$), and 1.96 ($R=20\himpc$) 
to correct for the underestimation of the bulk flow due to the limited 
size of the simulation box (see \S~\ref{theory_section}). 
In all cases, the standard deviation of the mean ($\sigma/\sqrt{N}$) 
is $\ltsim 0.04$ if one were to assume a Gaussian distribution. 
However, we show in \S~\ref{distribution_section} that the distribution is not 
well described by a Gaussian. 
For $R=10$ and $20\himpc$ case, the uncertainty is dominated by the cosmic variance; 
\ie, the number of independent spheres which fit in the simulation box.
See \S~\ref{method_section} for discussion. }

\end{deluxetable}

\clearpage

\stepcounter{thetabs}
\begin{deluxetable}{lccccccccccc}  
\tablenum{\tabnum}
\tablecolumns{12}  
\tablewidth{0pc}  
\tablecaption{Summary of the Bulk Flow in the Simulation\label{table2}}  
\tablehead{
\colhead{} & 
\multicolumn{3}{c}{$R=5\himpc$}  & \colhead{} &
\multicolumn{3}{c}{$R=10\himpc$} & \colhead{} &
\multicolumn{3}{c}{$R=20\himpc$} \\  
\cline{2-4} \cline{6-8} \cline{10-12} 
\colhead{} &  
\colhead{$\expec{V}$} & \colhead{$\expec{V^2}^{\frac{1}{2}}$} & \colhead{SD} & \colhead{} & 
\colhead{$\expec{V}$} & \colhead{$\expec{V^2}^{\frac{1}{2}}$} & \colhead{SD} & \colhead{} & 
\colhead{$\expec{V}$} & \colhead{$\expec{V^2}^{\frac{1}{2}}$} & \colhead{SD}}
\startdata 
\sidehead{Particle-based}
a) gal-pt  & 425 & 480 & 157 & & 384 & 432 & 127 & & 343 & 380 & 84 \\
b) dm-galctr-pt   & 419 & 469 & 148 & & 393 & 437 & 121 & & 370 & 404 & 82 \\
c) dm-dmctr-pt   & 472 & 528 & 165 & & 435 & 480 & 131 & & 394 & 427 & 85 \\
\tableline
\sidehead{Group-based}
a) gal-gp & 415 & 470 & 155 & & 368 & 417 & 124 & & 325 & 357 & 76 \\
b) dm-gp   & 493 & 551 & 172 & & 451 & 496 & 133 & & 408 & 439 & 83 \\
\enddata  

\tablecomments{
Mean and the rms value of the bulk flow in the 
simulation is summarized. SD stands for standard deviation.
All numbers are in units of $\kms$.
Each case corresponds to those in Table~\ref{table1}.
All numbers except the SD are after the multiplication by the
factors of 1.43 ($R=5\himpc$), 1.56 ($R=10\himpc$), and 1.96 ($R=20\himpc$) 
to correct for the underestimation due to the limited size of 
the simulation box (see \S~\ref{theory_section}). 
Discussions are in \S~\ref{mean_section}.}
\end{deluxetable}

\clearpage

\stepcounter{thetabs}
\begin{deluxetable}{lccccccccccc}  
\tablenum{\tabnum}
\tablecolumns{12}  
\tablewidth{0pc}  
\tablecaption{Summary of the Velocity Dispersion in the Simulation\label{table3}}  
\tablehead{
\colhead{} & 
\multicolumn{3}{c}{$R=5\himpc$}  & \colhead{} &
\multicolumn{3}{c}{$R=10\himpc$} & \colhead{} &
\multicolumn{3}{c}{$R=20\himpc$} \\  
\cline{2-4} \cline{6-8} \cline{10-12} 
\colhead{} &  
\colhead{$\expec{\sigma}$} & \colhead{$\expec{\sigma^2}^{\frac{1}{2}}$} & 
\colhead{SD} & \colhead{} & 
\colhead{$\expec{\sigma}$} & \colhead{$\expec{\sigma^2}^{\frac{1}{2}}$} & 
\colhead{SD} & \colhead{} & 
\colhead{$\expec{\sigma}$} & \colhead{$\expec{\sigma^2}^{\frac{1}{2}}$} & \colhead{SD}}
\startdata 
\sidehead{Particle-based}
a) gal-pt  & 289 & 342 & 182 & & 404 & 454 & 208 & & 521 & 553 & 184 \\
b) dm-galctr-pt   & 356 & 410 & 204 & & 432 & 483 & 215 & & 533 & 571 & 207 \\
c) dm-dmctr-pt   & 369 & 455 & 267 & & 465 & 538 & 271 & & 572 & 620 & 239 \\
\tableline
\sidehead{Group-based}
a) gal-gp & 274 & 329 & 182 & & 349 & 398 & 190 & & 434 & 466 & 171 \\
b) dm-gp   & 263 & 322 & 186 & & 339 & 385 & 183 & & 415 & 441 & 149 \\
\enddata  

\tablecomments{Mean and the rms value of the velocity dispersion 
$\sigma$ in the simulation is summarized. SD stands for standard deviation.
All numbers are in units of $\kms$.
Each case corresponds to those in Table~\ref{table1}.
See \S~\ref{mean_section} for discussion.}
\end{deluxetable}

\clearpage

\stepcounter{thetabs}
\begin{deluxetable}{ccccccccc}  
\tablenum{\tabnum}
\tablecolumns{9}  
\tablewidth{0pc}  
\tablecaption{Correlation between Galaxy Age and Overdensity\label{table4}}  
\tablehead{
\colhead{} & \colhead{} & \multicolumn{3}{c}{DWARF}   & \colhead{}
           & \multicolumn{3}{c}{GIANT} \\
\cline{3-5} \cline{7-9} \\
\colhead{} & \colhead{quartile} & 
\colhead{$\delta_{gal}$} & \colhead{$\delta_{L_V}$} & \colhead{$\delta_{DM}$} &  
\colhead{} & 
\colhead{$\delta_{gal}$} & \colhead{$\delta_{L_V}$} & \colhead{$\delta_{DM}$}}
\startdata
young & 1st & 3.45 & 3.21 & 2.41 &  & 0.81 & 1.37 & 0.43 \\
      & 2nd & 4.73 & 4.13 & 3.44 &  & 1.48 & 1.94 & 0.75 \\
      & 3rd & 4.69 & 4.11 & 3.30 &  & 2.32 & 2.53 & 1.25 \\
old   & 4th & 4.91 & 4.16 & 3.67 &  & 5.73 & 4.85 & 4.00 \\
\enddata
\tablecomments{Shown are the mean of the local overdensity for 
each quartile of galaxy sample divided in terms of its age.
Overdensity was calculated with a tophat $R=5\hinv$Mpc filter. 
Both $\delta_{gal}$ and $\delta_{DM}$ were calculated 
in terms of their mass, and $\delta_{L_V}$ is the luminosity-overdensity
calculated with absolute V-band luminosity. 
See \S~\ref{age_section} for discussion.}
\end{deluxetable}

\clearpage
\stepcounter{thetabs}
\begin{deluxetable}{cccccccccccc}  
\tablenum{\tabnum}
\tablecolumns{12}  
\tablewidth{0pc}  
\tablecaption{Galaxy Age Dependence of Mach Number\label{table5}}  
\tablehead{  
\colhead{} & \multicolumn{3}{c}{R=$5\hinv$Mpc}   & \colhead{}
           & \multicolumn{3}{c}{R=$10\hinv$Mpc} & \colhead{}
           & \multicolumn{3}{c}{R=$20\hinv$Mpc} \\  
\cline{2-4} \cline{6-8} \cline{10-12} 
\colhead{} & \colhead{median} & \colhead{mean} & \colhead{SDOM} &  
\colhead{} & \colhead{median} & \colhead{mean} & \colhead{SDOM} &
\colhead{} & \colhead{median} & \colhead{mean} & \colhead{SDOM}} 
\startdata 
\sidehead{DWARF}
young & 1.63 & 3.00 & 0.14 & & 1.11 & 1.44 & 0.06 & & 0.74 & 0.78 & 0.04 \\  
old   & 1.53 & 3.00 & 0.15 & & 1.19 & 1.54 & 0.06 & & 0.82 & 0.88 & 0.04 \\
\tableline
\sidehead{GIANT}
young & 2.13 & 2.86 & 0.06 & & 1.50 & 1.83 & 0.06 & & 0.98 & 1.04 & 0.04 \\
old   & 1.70 & 2.45 & 0.07 & & 1.20 & 1.53 & 0.06 & & 0.90 & 0.92 & 0.04 \\
\enddata  
\tablecomments{Shown are the mean, the median and the standard deviation 
of the mean (SDOM; $\sigma/\sqrt{N}$) of the Mach number for different
populations and scales. For $R=10$ and 20$\himpc$ cases, SDOM is limited 
by the independent number of spheres which fit in the simulation box.  
The values above are after the correction for the underestimation of 
the bulk flow due to the lack of long wavelength perturbation in the 
simulation. See \S~\ref{age_section} for discussion.}
\end{deluxetable}




\clearpage




\end{document}